\documentclass[aps,pre,reprint,superscriptaddress,longbibliography]{revtex4-2}
\usepackage{graphicx}
\usepackage{amsmath,amssymb}

\newcommand{\fr}[2]{\frac{\displaystyle #1}{\displaystyle #2}}
\newcommand{\df}[2]{\frac{\displaystyle d#1}{\displaystyle d#2}}

\newcommand{\pf}[2]{\frac{\displaystyle \partial #1}{\displaystyle \partial #2}}

\newcommand{\Real}{\mathrm{Re}\:}
\newcommand{\Imag}{\mathrm{Im}\:}

\newcommand{\qed}{\hfill \rule{2.3mm}{2.3mm}}

\newtheorem{proposition}{Proposition}
\newtheorem{remark}{Remark}

\newcommand{\be}{\begin{equation}}
\newcommand{\ee}{\end{equation}}

\begin{document}

\title{Periodic solutions in next generation neural field models}

\author{Carlo R. Laing}
\email{c.r.laing@massey.ac.nz}
\affiliation{
School of Mathematical and Computational Sciences, 
Massey University, Private Bag 102-904 NSMC, Auckland, New Zealand
}

\author{Oleh E. Omel'chenko}
\email{oleh.omelchenko@uni-potsdam.de}
\affiliation{
Institute of Physics and Astronomy, University of Potsdam, Karl-Liebknecht-Str. 24/25,
14476 Potsdam, Germany.
}


\date{\today}

\begin{abstract}
We consider a next generation neural field model
which describes the dynamics of a network of theta neurons on a ring.
For some parameters the network supports stable time-periodic solutions.
Using the fact that the dynamics at each spatial location are described
by a complex-valued Riccati equation we derive a self-consistency equation
that such periodic solutions must satisfy.
We determine the stability of these solutions,
and present numerical results to illustrate the usefulness of this technique.
The generality of this approach is demonstrated
through its application to several other systems involving delays,
two-population architecture and networks of Winfree oscillators.
\end{abstract}

\maketitle

%



\section*{Introduction}
The collective behaviour of large networks of neurons is a topic of ongoing interest.
One of the simplest forms of behaviour is
a periodic oscillation, which manifests itself
as a macroscopic rhythm created by the synchronous firing of many neurons.
Such oscillations have relevance to rhythmic movement~\cite{linpet22}, epilepsy~\cite{jirde13,netsch02},
schizophrenia~\cite{uhlsin10} neural communication~\cite{reyhug22} 
and EEG/MEG modelling~\cite{byrros22}, among others.
In different networks, oscillations may arise from mechanisms such as 
synaptic delays~\cite{devmon18},
the interaction of excitatory and inhibitory populations~\cite{borkop05,schavi20},
having sufficient connectivity in an inhibitory network~\cite{voltor18},
or the finite width of synaptic pulses emitted by neurons~\cite{ratpyr16}.
The modelling and simulation of such networks is essential in order to investigate their dynamics.

Among the many types of model neurons used when studying networks of neurons, theta
neurons~\cite{ermkop86}, Winfree oscillators~\cite{aristr01}
and quadratic integrate-and-fire (QIF) neurons~\cite{latric00} are some
of the simplest. These three types of model neurons have the advantage that
their mathematical form often allows infinite networks of such neurons to be analysed
exactly using the Ott-Antonsen method~\cite{ottant08,ottant09}.
We continue along those lines in this paper.

Here we largely consider a spatially-extended network of neurons, in which the neurons can be
thought of as lying on a ring. Such ring networks have been studied in connection with
modelling head direction~\cite{zha96} and working memory~\cite{funbru89,wimnyk14}, for example, and been studied by others~\cite{esnrox17,laicho01,kilerm13}.
We consider a network of theta neurons. The theta neuron is a minimal model for a neuron
which undergoes a saddle-node-on-invariant-circle (SNIC) bifurcation as a parameter is 
varied~\cite{ermkop86}. The theta neuron is exactly equivalent to a quadratic integrate-and-fire
(QIF) neuron, under the assumption of infinite firing threshold 
and reset values~\cite{monpaz15,avides22,devrox17}.
The coupling in the network is nonlocal synaptic coupling, implemented using a spatial
convolution with a translationally-invariant coupling kernel.

We studied such a model in the past~\cite{omelai22}, concentrating on describing 
spatially-uniform states and also stationary ``bump'' states in which there is a 
spatially-localised group of active neurons while the remainder of the network is
quiescent. We determined the existence and stability of such states and also found
regions of parameter space in which neither of these types of states were stable.
Instead, we sometimes found solutions which were periodic in time. In this paper we
study such periodic solutions using a recently-developed technique~\cite{ome23,ome22} 
which
is significantly faster than the standard approach. 

This technique was successfully applied to describe
traveling and breathing chimera states in nonlocally coupled phase oscillators~\cite{ome23,ome22}.
In this paper, we generalize this technique to neural models
and illustrate its possibilities with several examples.
We start with a ring network of theta neurons
and consider periodic bump states there.
We describe a continuation algorithm, perform linear stability analysis of such states
and derive some useful formulas, for example, for average firing rates.
We show that the same approach works in the presence of delays
and for two-population models. Finally, we show that Winfree oscillators
are also treatable by the proposed technique.

The structure of the paper is as follows.
In Sec.~\ref{Sec:Model} we present the discrete network model
and its continuum-level description.
The bulk of the paper is in Sec~\ref{sec:periodic},
where we show how to describe periodic states in a self-consistent way,
and show numerical results from implementing our algorithms.
Other models are considered in Sec.~\ref{sec:other} and we end in Sec.~\ref{sec:disc}.
The Appendix contains useful results regarding the complex Riccati equation.

\section{Theta neuron network model}
\label{Sec:Model}

The model we consider first is that in~\cite{omelai22,lai14a,laiome20}, which we briefly present here.
The discrete network consists of $N$ synaptically coupled theta neurons described by
\begin{equation}
\df{\theta_j}{t} = 1 - \cos\theta_j + ( 1 + \cos\theta_j ) ( \eta_j + \kappa I_j ),\quad j=1,\dots,N,
\label{Eq:Theta}
\end{equation}
where each $\theta_j\in[0,2\pi]$ is an angular variable. The constant $\kappa$ is the
overall strength of coupling within the network, and
the current entering the $j$th neuron is $\kappa I_j$ where
\begin{equation}
I_j(t) = \fr{2\pi}{N} \sum\limits_{k=1}^N K_{jk} P_n(\theta_k(t))
\label{eq:I}
\end{equation}
where
$$
P_n(\theta) = a_n ( 1 - \cos\theta )^n
$$
is a pulsatile function with a maximum at $\theta=\pi$ (when the neuron fires)
and $a_n$ is chosen according to the normalization condition
$$
\int_0^{2\pi} P_n(\theta) d\theta = 2\pi.
$$
Increasing $n$ makes $P_n(\theta)$ ``sharper'' and more pulse-like.
The excitability parameters $\eta_j$ are chosen from a Lorentzian distribution
with mean $\eta_0$ and width $\gamma > 0$
$$
g(\eta) = \fr{\gamma}{\pi} \fr{1}{( \eta - \eta_0 )^2 + \gamma^2}.
$$
The connectivity within the network is given by the weights $K_{jk}$ which are defined
by $K_{jk}=K(2\pi(j-k)/N)$ where the coupling kernel is
\begin{equation}
K(x) = \fr{1}{2\pi} ( 1 + A \cos x )
\label{Coupling:Cos}
\end{equation}
for some constant $A$. Note that the form of coupling implies that the neurons
are equally-spaced around a ring, with periodic boundary conditions.
Such a network can support solutions which are --- at a macroscopic level
--- periodic in time; see Fig.~3 in~\cite{omelai22}.
Such solutions are unlikely to be true periodic solutions of~\eqref{Eq:Theta}, 
since for a typical realisation of the $\eta_j$,
one or more neurons will have extreme values of this parameter,
resulting in them not frequency-locking to other neurons.

Note that the model we study here has only one neuron at each spatial position,
and for $A>0$ the connections between nearby neurons are more positive
than those between distant neurons.
For $A>1$ neurons on opposite sides of the domain inhibit one another, as the connections
between them are negative, giving a ``Mexican-hat'' connectivity. For $A<0$ connections
between neurons on opposite sides of the domain are more positive than those between
nearby neurons. Such a model with
one population of neurons and connections of mixed sign can be thought of as
an approximation of a network
with populations of both excitatory and inhibitory neurons~\cite{pinerm01,esnrox17}.


Using the Ott/Antonsen ansatz~\cite{ottant08,ottant09},
one can show that in the limit $N\to\infty$,
the long-term dynamics of the network~\eqref{Eq:Theta} can be described by
\begin{eqnarray}
\pf{z}{t} & = & \fr{(i \eta_0 - \gamma) ( 1 + z )^2 - i ( 1- z )^2 }{2} \nonumber \\ 
& + & \kappa \fr{ i ( 1 + z )^2 }{2} \mathcal{K} H_n(z),
\label{Eq:L}
\end{eqnarray}
where
\begin{equation}
(\mathcal{K} \varphi) (x) = \int_0^{2\pi} K(x - y) \varphi(y) dy
\label{Def:K}
\end{equation}
is the convolution of $K$ and $\varphi$
and
$$
H_n(z) = a_n \left[ C_0 + \sum\limits_{q=1}^n C_q \left( z^q + \overline{z}^q \right) \right],
$$
where
$$
C_q = \sum\limits_{k=0}^n \sum\limits_{m=0}^k \fr{\delta_{k-2m,q} (-1)^k n!}{2^k (n - k)! m! (k - m)!}.
$$
For our computations we set $n=2$, so that
\[
   H_2(z) =(2/3)[3/2-(z+\bar{z})+(z^2+\bar{z}^2)/4]
\]
where overline indicates the complex conjugate.
Periodic solutions like those studied below were also found with $n=5$,
for example, so the choice of $n$ is not critical.
The wider question of the effects of pulse shape
and duration is an interesting one~\cite{pie23}.

Eq.~\eqref{Eq:L} is an integro-differential equation for a complex-valued function
$z(x,t)$, where $x\in[0,2\pi]$ is position on the ring. $z(x,t)$ is a local order parameter
and can be thought of as the average of $e^{i\theta}$ for neurons in a small neighbourhood
of position $x$. The magnitude of $z$ is a measure of how synchronised the neurons are,
whereas its argument gives the most likely value of $\theta$~\cite{lai14a}.
Using the equivalence of theta and QIF neurons,
one can also provide a relevant biological interpretation of $z$.
Namely, defining $W \equiv (1 - \overline{z})/(1 + \overline{z})$,
one can show~\cite{lai15,monpaz15} that $\pi^{-1} \Real W$ is the flux through $\theta = \pi$
or the instantaneous firing rate of neurons at position~$x$ and time~$t$.
Similarly, if $V_j = \tan(\theta_j/2)$ is the voltage of the $j$th QIF neuron,
the mean voltage at position $x$ and time $t$ is given by $\Imag W$.
Note that physically relevant solutions of Eq.~(\ref{Eq:L}) must assume values $|z|\le 1$.
In other words, we are interested only in solutions $z\in\overline{\mathbb{D}}$, where
$$
\mathbb{D} = \{ z\in\mathbb{C}\::\: |z| < 1 \}
$$
is the unit disc in the complex plane.

Equations of the form~\eqref{Eq:L}--\eqref{Def:K} are sometimes referred to as
{\it next generation neural field models}~\cite{byravi19,coobyr19} as they have the form of a
neural field model (an integro-differential equation for a macroscopic quantity such as
``activity''~\cite{laitro02,ama77}) but are derived exactly from a network like~\eqref{Eq:Theta},
rather than being of a phenomenological nature.

\section{Periodic states}
\label{sec:periodic}

\begin{figure}[b]
\centering
\includegraphics[scale=0.8]{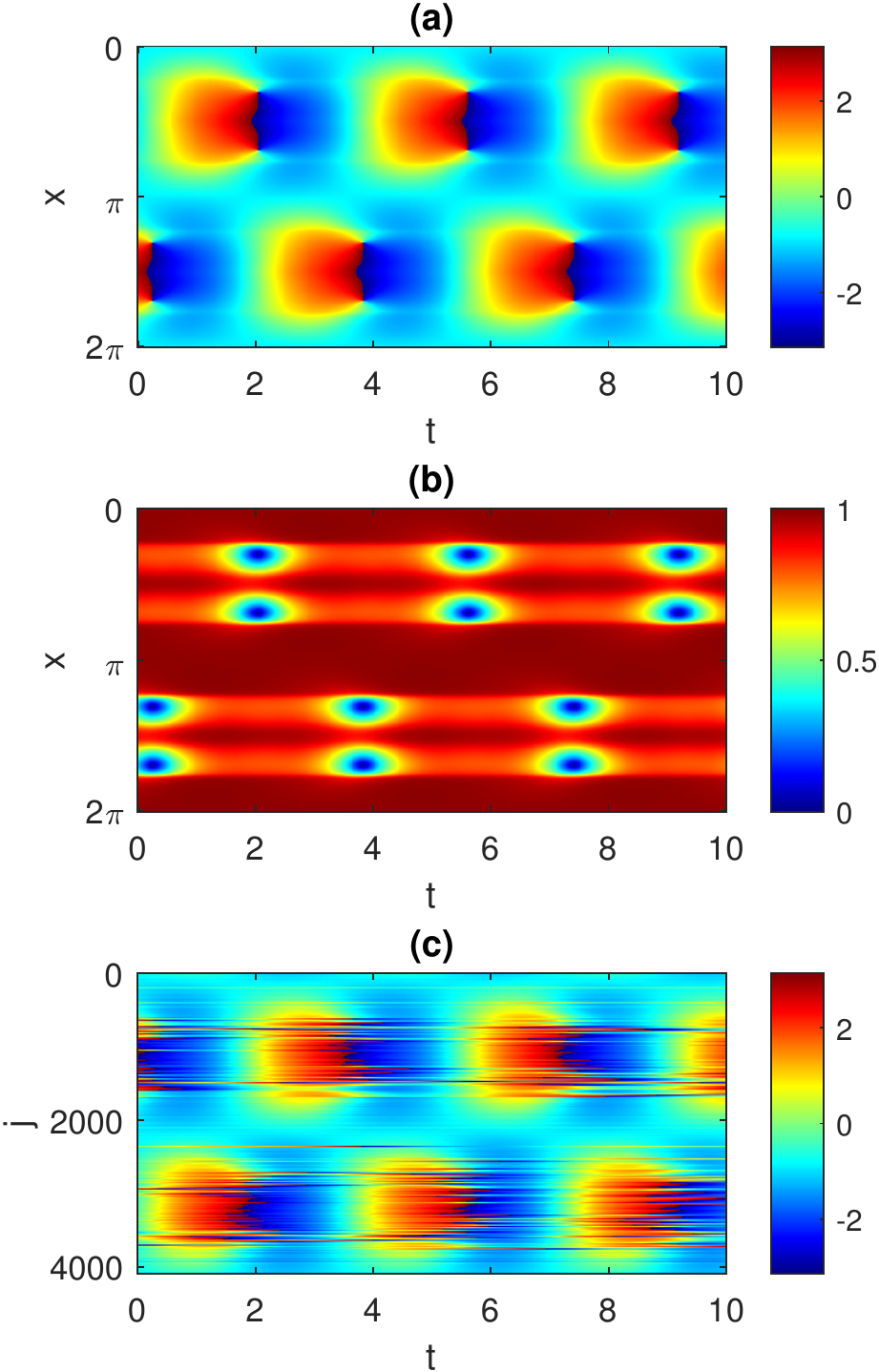}
\caption{A typical periodic solution of Eq.~\eqref{Eq:L}. (a): $\arg{(z(x,t))}$. (b): $|z(x,t)|$.
(c) A realization of this solution in a network
of $N = 4096$ theta neurons described by Eq.~(\ref{Eq:Theta}).
$\theta_j$ is shown in color.
Parameters: $A=-5$, $\eta_0=-0.7$, $\kappa=1$, $\gamma=0.01$.}
\label{fig:symm}
\end{figure}

In this paper we focus on states with periodically oscillating macroscopic dynamics.
For the mean field equation~(\ref{Eq:L}) these correspond to periodic solutions
$$
z = a(x,t)
$$
where $a(x,t+T) = a(x,t)$ for some $T > 0$.
The frequency corresponding to $T$ is denoted by $\omega = 2\pi / T$.
An example of a periodic solution is shown in Fig.~\ref{fig:symm}, panels~(a) and~(b).
Fig.~\ref{fig:symm}~(c) shows a realization of the same solution in the discrete 
network~\eqref{Eq:Theta}.
We note that this solution has a spatio-temporal symmetry:
it is invariant under a time shift of half a period followed by a reflection about $x=\pi$.
However, we do not make use of this symmetry in the following calculations.

The appearance of periodic solutions in this model is in contrast with classical 
one-population neural field models for which they do not seem to occur~\cite{lai03}.
This is another example of next generation neural field models showing more complex
time-dependent behaviour than classical ones~\cite{laiome20}.

A straight-forward way to study a periodic orbit like that in Fig.~\ref{fig:symm}
would be discretise Eq.~\eqref{Eq:L} on a spatially-uniform grid
and approximate the convolution using matrix/vector multiplication or otherwise,
resulting in a large set of coupled ordinary differential equations.
The periodic solution of these could then be studied using standard techniques~\cite{lai14a},
but note that the computational complexity of this would typically scale as $\sim N^2$,
where $N$ is the number of spatial points used in the grid.
Instead we propose here an alternative method based on the ideas from~\cite{ome23,ome22},
which allows us to perform the same calculations with only $\sim N$ operations.
The main ingredients of this method are explained in Sections~\ref{Sec:Moebius} and~\ref{Sec:SC}.
They include the description of the properties of complex Riccati equation
and the derivation of the self-consistency equation for periodic solutions of Eq.~(\ref{Eq:L}).
In Section~\ref{Sec:Numerics} we explain how the self-consistency equation
can be solved in the case of coupling function~(\ref{Coupling:Cos}).
Then in Section~\ref{Sec:Results} we report some numerical results obtained with our method.
In addition, in Section~\ref{Sec:Stability} we perform
a rigorous linear stability analysis of periodic solutions of Eq.~(\ref{Eq:L})
by considering the spectrum of the corresponding monodromy operator.
Finally, in Section~\ref{Sec:FiringRates}, we show how the mean field equation~\eqref{Eq:L}
can be used to predict the average firing rate distribution in the neural network~\eqref{Eq:Theta}.

\subsection{Periodic complex Riccati equation and M{\"o}bius transformation}
\label{Sec:Moebius}

By the time rescaling $u(x,t) = z(x,t/\omega)$ we can rewrite Eq.~(\ref{Eq:L}) in the form
\begin{eqnarray}
\omega \pf{u}{t} & = & \fr{(i \eta_0 - \gamma) ( 1 + u )^2 - i ( 1- u )^2 }{2} \nonumber \\
& + & \kappa \fr{ i ( 1 + u )^2 }{2} \mathcal{K} H_n(u),
\label{Eq:L_}
\end{eqnarray}
such that the above $T$-periodic solution of Eq.~(\ref{Eq:L})
corresponds to a $2\pi$-periodic solution of Eq.~(\ref{Eq:L_}).
Then, dividing~(\ref{Eq:L_}) by~$\omega$
and reordering the terms we can see that Eq.~(\ref{Eq:L_})
is equivalent to a complex Riccati equation
\begin{eqnarray}
\pf{u}{t} &=& i \left( W(x,t) + \zeta - \fr{1}{2\omega} \right) + 2 i \left( W(x,t) + \zeta + \fr{1}{2\omega} \right) u \nonumber \\
& + & i \left( W(x,t) + \zeta - \fr{1}{2\omega} \right) u^2,
\label{Eq:Riccati}
\end{eqnarray}
with the $2\pi$-periodic in~$t$ coefficient
\begin{equation}
W(x,t) = \fr{\kappa}{2\omega} \mathcal{K} H_n(u)
\label{Def:W}
\end{equation}
and
\begin{equation}
\zeta = \fr{\eta_0 + i \gamma}{2\omega}.
\label{Def:zeta}
\end{equation}
In~\cite{ome23} it was shown
(see also Proposition~\ref{Proposition:Sln:Theta}
and Remark~\ref{Remark:Sln:Theta} in the Appendix for additional details)
that independent of the choice of the real-valued periodic function $W(x,t)$,
parameters $\zeta\in\mathbb{C}_\mathrm{up}=\{ z\in\mathbb{C}\::\: \Imag z > 0\}$ and $\omega > 0$,
for every fixed $x\in[0,2\pi]$ Eq.~(\ref{Eq:Riccati}) has a unique
stable $2\pi$-periodic solution $U(x,t)$
that lies entirely in the open unit disc~$\mathbb{D}$.
Denoting the corresponding solution operator by
$$
\mathcal{U}\::\: C_\mathrm{per}([0,2\pi];\mathbb{R})\times\mathbb{C}_\mathrm{up}\times(0,\infty)\to C_\mathrm{per}([0,2\pi];\mathbb{D}),
$$
we can write the $2\pi$-periodic solution of interest as
\begin{equation}
U(x,t) = \mathcal{U}\left( W(x,t), \fr{ \eta_0 + i \gamma }{2\omega}, \omega \right).
\label{Formula:W_to_u}
\end{equation}
Note that $C_\mathrm{per}([0,2\pi];\mathbb{R})$ here denotes
the space of all real-valued continuous $2\pi$-periodic functions,
while the notation $C_\mathrm{per}([0,2\pi];\mathbb{D})$ stands for
the space of all complex continuous $2\pi$-periodic functions
with values in the open unit disc~$\mathbb{D}$.
Importantly, the variable~$x$ appears in formula~(\ref{Formula:W_to_u}) as a parameter
so that the function $W(x,\cdot)\in C_\mathrm{per}([0,2\pi];\mathbb{R})$ with a fixed $x$
is considered as the first argument of the operator~$\mathcal{U}$.

As for the operator $\mathcal{U}$, although it is not explicitly given,
its value can be calculated without resource-demanding iterative methods
by solving exactly four initial value problems for Eq.~(\ref{Eq:Riccati}).
The rationale for this approach can be found in~\cite[Section~4]{ome23}
and is repeated for completeness in Remark~\ref{Remark:Sln:Operator} in Appendix.
Below we describe its concrete implementation in the case of formula~(\ref{Formula:W_to_u}).

We assume that the spatial domain $[0,2\pi]$ is discretised with $N$ points, $x_j, j=1,2,\dots N$
and that the functions $U(x,t)$ and $W(x,t)$ are replaced with their grid counterparts
$u_j(t) = U(x_j,t)$ and $w_j(t) = W(x_j,t)$, respectively.
Then, given a set of functions $w_j(t)$,
we calculate the corresponding functions $u_j(t)$
by performing the following four steps.

(i) We (somewhat arbitrarily) choose three initial conditions
$u_j^1(0)=-0.95$, $u_j^2(0)=0$, $u_j^3(0)=0.95$
and solve Eq.~\eqref{Eq:Riccati} with these initial conditions
to obtain solutions $u_j^k(t)$, $j=1,\dots N$; $k=1,2,3$.
In~\cite{ome23}
(see also Proposition~\ref{Proposition:Sln:Theta} in the Appendix)
it is shown that for $\gamma > 0$
these solutions lie in the open unit disc $\mathbb{D}$.

(ii) Since at each point in space the Poincar{\'e} map of Eq.~(\ref{Eq:Riccati}) with $2\pi$-periodic coefficients
coincides with a M{\"o}bius map~$\mathcal{M}(u)$ (see~\cite{ome23} for detail),
we can use the relations $u_j^k(2\pi) = \mathcal{M}_j( u_j^k(0) )$, $k=1,2,3$,
to reconstruct these maps $\mathcal{M}_j$, $j=1,\dots N$. 
The corresponding formulas are given in~\cite[Section~4]{ome23}.

(iii) Now that the M{\"o}bius maps $\mathcal{M}_j(u)$ are known,
their fixed points can be found by solving $u_j^\ast = \mathcal{M}_j(u_j^\ast)$ for each $j$.
This equation is equivalent to a complex quadratic equation,
therefore in general it has two solutions in the complex plane.
For $\gamma > 0$, only one of these solutions lies in the unit disc $\mathbb{D}$.

(iv) Using the fixed points $u_j^\ast\in\mathbb{D}$ of the M{\"o}bius maps $\mathcal{M}_j(u)$
as an initial condition in Eq.~(\ref{Eq:Riccati}) (i.e.~setting $u(x_j,0)=u_j^\ast$) 
and integrating~\eqref{Eq:Riccati} for a fourth time to $t=2\pi$
we obtain the grid counterpart of a $2\pi$-periodic solution, $U(x,t)$,
that lies entirely in the unit disc $\mathbb{D}$.

We now use this result to show how to derive a self-consistency equation,
the solution of which allows us to determine a $2\pi$-periodic solution of Eq.~\eqref{Eq:L_}.

\subsection{Self-consistency equation}
\label{Sec:SC}

Supposing that Eq.~(\ref{Eq:L_}) has a $2\pi$-periodic solution,
then using formula~(\ref{Def:W}) we can calculate the corresponding function $W(x,t)$.
On the other hand, using formula~\eqref{Formula:W_to_u} we can recover $u(x,t)$.
Then the new and the old expressions of $u(x,t)$ will agree with each other
if and only if the function $W(x,t)$ satisfies a self-consistency equation
\begin{equation}
W(x,t) = \fr{\kappa}{2\omega} \mathcal{K} H_n\left( \mathcal{U}\left( W(x,t), \fr{\eta_0 + i \gamma}{2\omega}, \omega \right) \right),
\label{Eq:SC}
\end{equation}
obtained by inserting~\eqref{Formula:W_to_u} into~(\ref{Def:W}).
In the following we consider Eq.~(\ref{Eq:SC}) as a separate equation
which must be solved with respect to $W(x,t)$ and $\omega$.
Note that the unknown field $W(x,t)$ has a problem-specific meaning:
It is proportional to the current entering a neuron at position $x$ at time $t$
due to the activity of all other neurons in the network.
The use of self-consistency arguments to study infinite networks of oscillators
goes back to Kuramoto~\cite{kuramotobook84,str00,shikur04},
but such approaches have always focused on steady states,
whereas we consider periodic solutions here.

Note that from a computational point of view,
the self-consistency equation~(\ref{Eq:SC}) has many advantages.
It allows us to reduce the dimensionality of the problem
at least in the case of special coupling kernels
with finite number of Fourier harmonics (see Section~\ref{Sec:Numerics}).
Moreover, its structure is convenient for parallelization,
since the computations of operator $\mathcal{U}$ at different points $x$
are performed independently.
Finally, the main efficiency is due to the fact that the computation of $\mathcal{U}$
is performed in non-iterative way.

In the next proposition, we will prove some properties
of the solutions of Eq.~(\ref{Eq:SC}), which will be used later in Section~\ref{Sec:Stability}.

\begin{proposition}
Let the pair $( W(x,t), \omega )$ be a solution of the self-consistency equation~(\ref{Eq:SC})
and let $U(x,t)$ be defined by~(\ref{Formula:W_to_u}). Then
$$
\left| \exp\left( \int_0^{2\pi} M(x,t) dt \right) \right| < 1
$$
where
\begin{eqnarray}
M(x,t) &=& \fr{i}{\omega} \left[ ( \kappa \mathcal{K} H_n(U) + \eta_0 + i \gamma ) ( 1 + U(x,t) )\right. \nonumber \\
& + & \left. 1 - U(x,t) \right].
\label{Def:M}
\end{eqnarray}
\label{Proposition:Stability}
\end{proposition}

{\bf Proof:} For every fixed $x\in[0,2\pi]$, the function $U(x,t)$ yields
a stable $2\pi$-periodic solution of the complex Riccati equation~(\ref{Eq:Riccati}).
The linearization of Eq.~(\ref{Eq:Riccati}) around this solution reads
$$
\df{v}{t} = M(x,t) v,
$$
where
\begin{eqnarray*}
M(x,t) &=& 2 i \left( W(x,t) + \zeta + \fr{1}{2\omega} \right) \nonumber \\
& + & 2 i \left( W(x,t) + \zeta - \fr{1}{2\omega} \right) U(x,t).
\end{eqnarray*}
Moreover, using~(\ref{Def:W}) and~(\ref{Def:zeta}),
we can show that the above expression determines
a function identical to the function $M(x,t)$ in~(\ref{Def:M}).
Recalling that $U(x,t)$ is not only a stable but also an asymptotically stable solution of Eq.~(\ref{Eq:Riccati}),
see Remark~\ref{Remark:AsymptoticStability} in Appendix, we conclude
that the corresponding Floquet multiplier lies in the open unit disc $\mathbb{D}$.
This ends the proof.~\qed

\subsection{Numerical implementation}
\label{Sec:Numerics}

Eq.~(\ref{Eq:SC}) describes a periodic orbit,
and since Eq.~\eqref{Eq:Riccati} is autonomous we need to append a pinning condition
in order to select a specific solution of Eq.~(\ref{Eq:SC}). For a solution of the type shown
in Fig.~\ref{fig:symm} we choose
\begin{equation}
\int_0^{2\pi} dx \int_0^{2\pi} W(x,t) \sin(2 t) dt = 0.
\label{Eq:Pinning}
\end{equation}

In the following we focus on the case of the cosine coupling~(\ref{Coupling:Cos}).
It is straight-forward to show that $(\mathcal{K} H_n)(x)$ can be written as a linear
combination of $1,\cos{x}$ and $\sin{x}$. However, the system is translationally invariant
in $x$, and we can
eliminate this invariance from Eq.~(\ref{Eq:SC}) by
restricting this equation to its invariant subspace $\mathrm{Span}\{ 1, \sin x \}$.
Then, taking into account that the function $W(x,t)$ is real,
we seek an approximate solution of the system~(\ref{Eq:SC}), (\ref{Eq:Pinning})
using a Fourier-Galerkin ansatz
\begin{equation}
W(x,t) = \sum\limits_{m=0}^{2F} ( v_m + w_m \sin x ) \psi_m(t)
\label{W:approx}
\end{equation}
where $v_m$ and $w_m$ are real coefficients
and $\psi_m(t)$ are trigonometric basis functions
\begin{eqnarray*}
\psi_0(t) &=& 1,\\[2mm]
\psi_m(t) &=& \sqrt{2} \cos(n t),\quad\mbox{if}\quad m = 2 n\quad\mbox{with}\quad n\in\mathbb{N},\\[2mm]
\psi_m(t) &=& \sqrt{2} \sin(n t),\quad\mbox{if}\quad m = 2 n -1\quad\mbox{with}\quad n\in\mathbb{N}.
\end{eqnarray*}
Our typical choice of the number of harmonics in~(\ref{W:approx}) is $F = 10$.
To exactly represent $W(x,t)$ in~\eqref{W:approx} would require an infinite
number of terms in the series, so using a finite value of $F$ introduces an approximation
in our calculations. However, the excellent agreement between our calculations with $F=10$
and those from full simulations of~\eqref{Eq:L} (shown below) indicate
that such an approximation is justified.

Using the scalar product
$$
\langle u, v \rangle = \fr{1}{(2\pi)^2} \int_0^{2\pi} dx \int_0^{2\pi} u(x,t) v(x,t) dt
$$
we project Eq.~(\ref{Eq:SC}) on different spatio-temporal Fourier modes to obtain the system
\begin{eqnarray}
v_m &=& \fr{\kappa}{2\omega} \left\langle H_n\left( \mathcal{U}\left( W(x,t), \fr{\eta_0 + i \gamma}{2\omega}, \omega \right) \right), \psi_m(t) \right\rangle,
\label{Eq:vm}\\[2mm]
w_m &=& \fr{\kappa A}{2\omega} \left\langle H_n\left( \mathcal{U}\left( W(x,t), \fr{\eta_0 + i \gamma}{2\omega}, \omega \right) \right), \right.\nonumber \\
&& \left. \vphantom{H_n\left( \mathcal{U}\left( W(x,t), \fr{\eta_0 + i \gamma}{2\omega}, \omega \right) \right)} \psi_m(t) \sin x \right\rangle,
\label{Eq:wm}
\end{eqnarray}
for $m=0,1,\dots, 2F$.
Eqs.~\eqref{Eq:vm} and~\eqref{Eq:wm}, together with \eqref{Eq:Pinning},
are a set of  $2(2F+1)+1=4F+3$ equations for the $4F+3$ unknowns 
$v_0,v_1,\dots, v_{2F},w_0,w_1,\dots, w_{2F},\omega$, which must be solved simultaneously.
We solve them using Newton's method and find convergence within 3 or 4 iterations.

In simple terms, suppose we have somewhat accurate estimates of 
$v_0,v_1,\dots, v_{2F},w_0,w_1,\dots, w_{2F},\omega$.
These can be inserted into~\eqref{W:approx}
to calculate the function $W(x,t)$. Then
one can calculate a periodic solution of Eq.~(\ref{Eq:Riccati}) with the specified $W(x,t)$
by formula~(\ref{Formula:W_to_u})
and finally calculate $H_n( \mathcal{U})$ and insert this into~\eqref{Eq:vm} and \eqref{Eq:wm}
and perform the projections. We want the difference between the new values of
$v_0,v_1,\dots, v_{2F},w_0,w_1,\dots, w_{2F}$ and the initial values of them to be zero,
and for~\eqref{Eq:Pinning} to hold. This determines the $4F+3$ equations we need to solve.
The solutions of these equations can be followed as a parameter is varied in the standard
way~\cite{lai14b}.
Note that these calculations involve discretising the spatial
domain with $N$ points. However, the number of unknowns ($4F+3$) is significantly less than $N$.

\subsection{Results}
\label{Sec:Results}

\begin{figure}[t]
\centering
\includegraphics[scale=0.7]{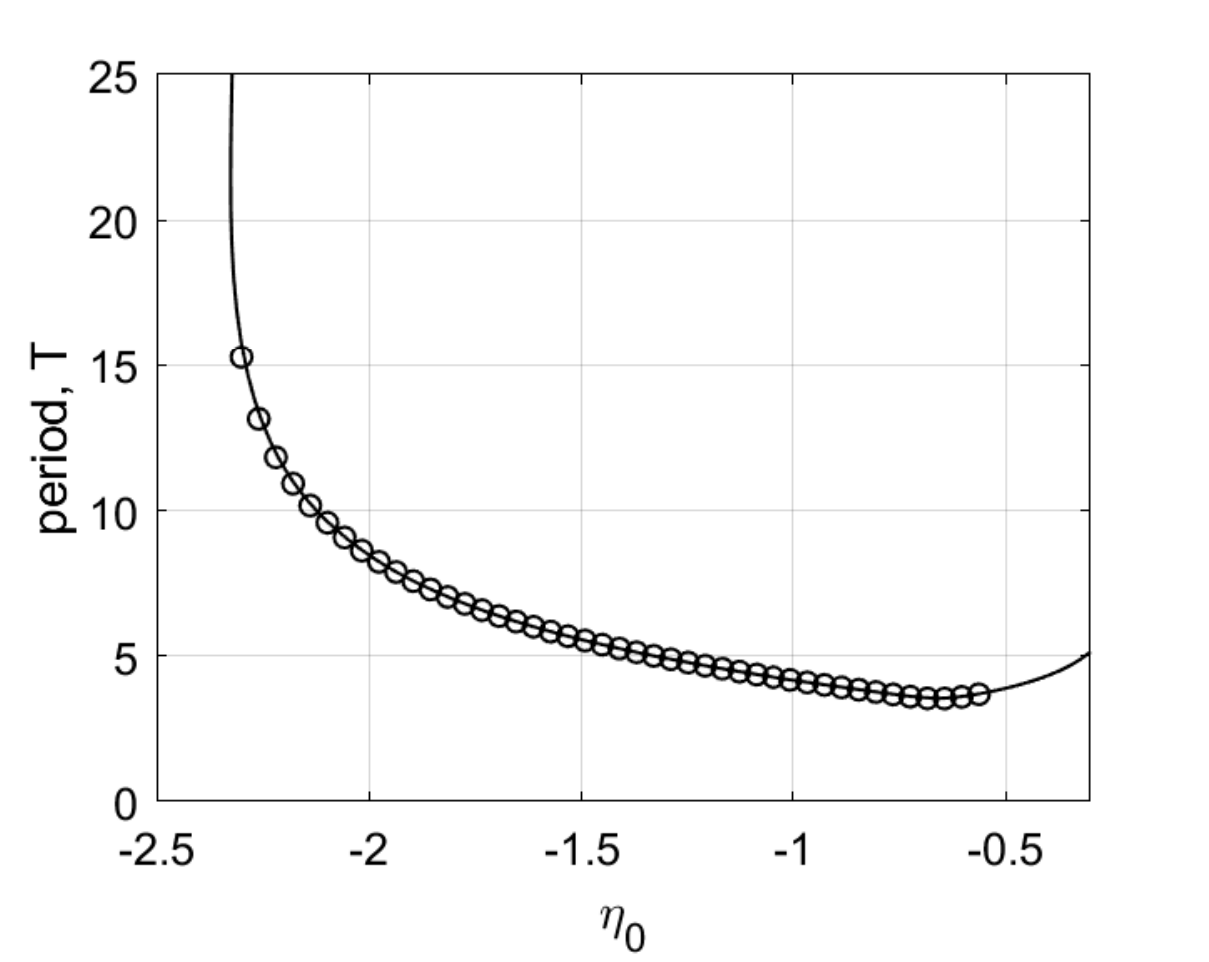}
\caption{Period of the type of solution shown in Fig.~\ref{fig:symm} as a function of $\eta_0$
(solid curve). The circles show values measured from direct simulations of Eq.~\eqref{Eq:L}.
Other parameters: $A=-5$, $\kappa=1$, $\gamma=0.01$.}
\label{fig:symmperiod}
\end{figure}

The results of following the solution shown in Fig.~\ref{fig:symm} as $\eta_0$ is varied
are shown in Fig.~\ref{fig:symmperiod},
along with values measured from direct simulations of Eq.~\eqref{Eq:L}.
The period seems to become arbitrarily large as $\eta_0$ approaches $-2.32$,
and the solution approaches a heteroclinic connection, spending more and more time near
two symmetric states which are mapped to one another under the transformation $x\to -x$.
The solution becomes unstable at $\eta_0\approx -0.5$ through a subcritical
torus (or secondary Hopf) bifurcation. (This was determined by finding the Floquet
multipliers of the periodic solution in the conventional way; results not shown.)
For these calculations we used $N=256$ spatial points.

\begin{figure}[t]
\centering
\includegraphics[scale=0.75]{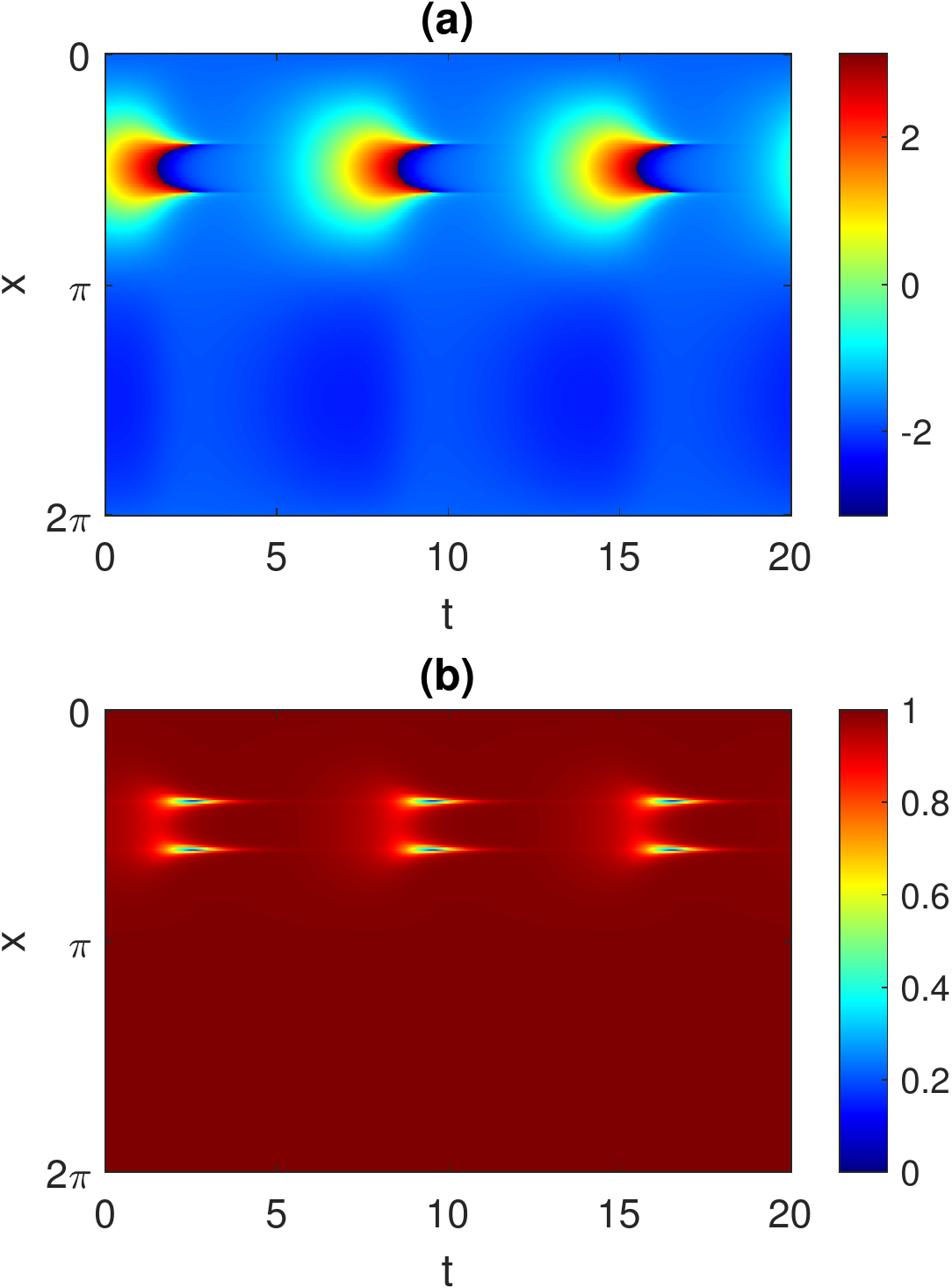}
\caption{Another periodic solution of Eq.~\eqref{Eq:L}. (a): $\arg{(z(x,t))}$. (b): $|z(x,t)|$.
Parameters: $A=-5$, $\eta_0=-2.5$, $\kappa=1$, $\gamma=0.01$.}
\label{fig:asymmA}
\end{figure}

\begin{figure}[h!]
\centering
\includegraphics[scale=0.55]{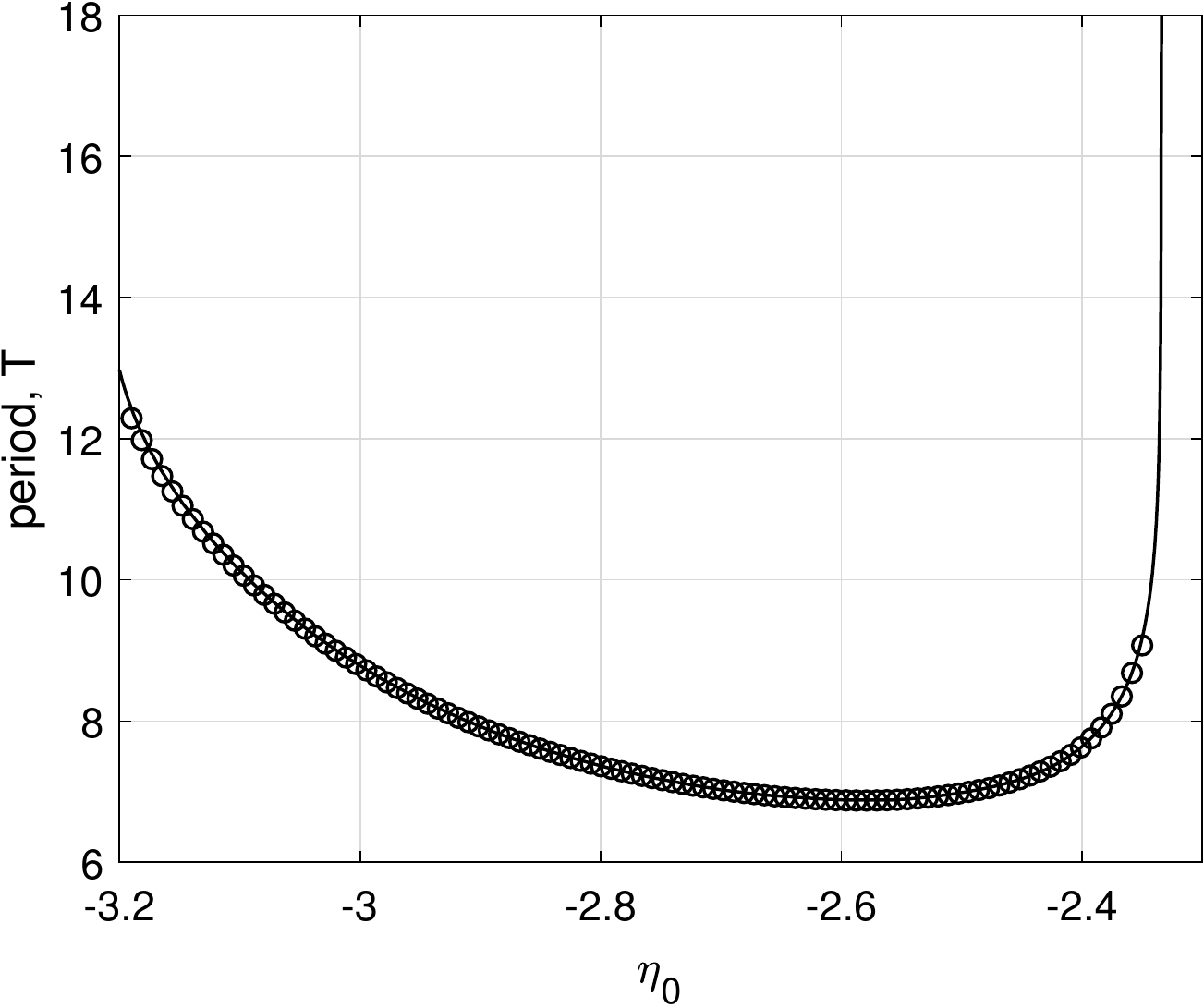}
\caption{Period of the type of solution shown in Fig.~\ref{fig:asymmA} as a function of $\eta_0$
(solid curve). The circles show values measured from direct simulations of Eq.~\eqref{Eq:L}.
Other parameters: $A=-5$, $\kappa=1$, $\gamma=0.01$.}
\label{fig:asymmperiod}
\end{figure}

For more negative values of $\eta_0$ than those shown in Fig.~\ref{fig:symmperiod}, another
type of periodic solution is stable: see Fig.~\ref{fig:asymmA}. Such a solution does not have
the spatio-temporal symmetry of the solution shown in Fig.~\ref{fig:symm}. However, we
can follow it in just the same way as the parameter $\eta_0$ is varied, and we obtain the 
results shown in Fig.~\ref{fig:asymmperiod}. This periodic orbit appears to be destroyed
in a supercritical Hopf bifurcation as $\eta_0$ is decreased through approximately $-3.2$,
and become unstable to a wandering pattern at $\eta_0$ is increased through  
approximately $-2.34$.

Note that the left asymptote in Fig.~\ref{fig:symmperiod} coincides with the
right asymptote in Fig.~\ref{fig:asymmperiod}. On
the other hand, we note that two patterns shown in Figs.~\ref{fig:symm} 
and~\ref{fig:asymmA} have different spatiotemporal
symmetries, therefore due to topological reasons they cannot continuously transform into each
other. Similar bifurcation diagrams where parameter ranges of two patterns with different
symmetries are separated by heteroclinic or homoclinic bifurcations were found for non-locally
coupled Kuramoto-type phase oscillators~\cite{ome20A} and seem to be a general
mechanism which, however, needs additional investigation.

\subsection{Stability of breathing bumps}
\label{Sec:Stability}

Given a $T$-periodic solution $a(x,t)$ of Eq.~(\ref{Eq:L}),
we can perform its linear stability analysis,
using the approach proposed in~\cite{ome22}. Before doing this, we write
\[
   H_n(z)=a_nC_0+2\mbox{Re}[D_n(z)]
\]
where
\[
   D_n(z)=a_n\sum_{q=1}^n C_qz^q,
\]
to emphasise that $H_n(z)$ is always real.
Now, we insert the ansatz $z(x,t) = a(x,t) + v(x,t)$ into Eq.~(\ref{Eq:L})
and linearize the resulting equation with respect to small perturbations $v(x,t)$.
Thus, we obtain a linear integro-differential equation
\begin{equation}
\pf{v}{t} = \mu(x,t) v  + \kappa \fr{ i ( 1 + a(x,t) )^2 }{2}
\mathcal{K} \left( D_n'(a) v + \overline{D_n'(a)} \overline{v} \right),
\label{Eq:v}
\end{equation}
where
\begin{equation}
\mu(x,t) = [ i (\eta_0 + \kappa \mathcal{K} H_n(a)) - \gamma ] ( 1 + a(x,t) ) + i ( 1- a(x,t) )
\label{Def:mu}
\end{equation}
and
$$
D_n'(z) = \df{}{z} D_n(z) = a_n \sum\limits_{q=1}^n q C_q z^{q-1}.
$$
Note that Eq.~(\ref{Eq:v}) coincides with the Eq.~(5.1) from~\cite{omelai22},
except that the coefficients $a(x,t)$ and $\mu(x,t)$ are now time-dependent.
Since Eq.~(\ref{Eq:v}) contains the complex-conjugated term~$\overline{v}$,
it is convenient to consider this equation along with its complex-conjugate 
$$
\pf{\overline{v}}{t} = \overline{\mu}(x,t) \overline{v}  - \kappa \fr{ i ( 1 + \overline{a}(x,t) )^2 }{2}
\mathcal{K} \left( D_n'(a) v + \overline{D_n'(a)} \overline{v} \right).
$$
This pair of equations can be written in the operator form
\begin{equation}
\df{V}{t} = \mathcal{A}(t) V + \mathcal{B}(t) V,
\label{Eq:V}
\end{equation}
where $V(t) = ( v_1(t), v_2(t) )^\mathrm{T}$ is a function
with values in $C_\mathrm{per}([0,2\pi]; \mathbb{C}^2)$, and
$$
\mathcal{A}(t) V = \left(
\begin{array}{cc}
\mu(x,t) & 0 \\[2mm]
0 & \overline{\mu}(x,t)
\end{array}
\right)
\left(
\begin{array}{c}
v_1 \\[2mm]
v_2
\end{array}
\right),
$$
and
\begin{equation*}
\mathcal{B}(t) V = \fr{i \kappa}{2} \left(
\begin{array}{cc}
( 1 + a(x,t) )^2 & 0 \\[2mm]
0 & - ( 1 + \overline{a}(x,t) )^2
\end{array}
\right)
\end{equation*}
\begin{equation}
\times \left(
\begin{array}{c}
\mathcal{K} \left( D_n'(a) v_1 \right) \\[2mm]
\mathcal{K} \left( \overline{D_n'(a)} v_2 \right)
\end{array}
\right).
\label{Def:B}
\end{equation}
For every fixed $t$ the operators $\mathcal{A}(t)$ and $\mathcal{B}(t)$
are linear operators from $C_\mathrm{per}([0,2\pi]; \mathbb{C}^2)$ into itself.
Moreover, they both depend continuously on $t$
and thus their norms are uniformly bounded for all $t\in[0,T]$.

Recall that the question of linear stability of $a(x,t)$ in Eq.~(\ref{Eq:L}) is equivalent
to the question of linear stability of the zero solution in Eq.~(\ref{Eq:v}),
and hence to the question of linear stability of the zero solution in Eq.~(\ref{Eq:V}).
Moreover, using the general theory of periodic differential equations in Banach spaces,
see \cite[Chapter~V]{DaleckiiKrein},
the last question can be reduced to the analysis of the spectrum
of the monodromy operator $\mathcal{E}(T)$ defined by the operator exponent
$$
\mathcal{E}(t) = \exp\left( \int_0^t ( \mathcal{A}(t') + \mathcal{B}(t') ) dt' \right).
$$
The analysis of Eq.~(\ref{Eq:V}) in the case when $\mathcal{A}(t)$
is a matrix multiplication operator and $\mathcal{B}(t)$
is an integral operator similar to~(\ref{Def:B})
has been performed in~\cite[Section~4]{ome22}.
Repeating the same arguments we can demonstrate
that the spectrum of the monodromy operator $\mathcal{E}(T)$
is bounded and symmetric with respect to the real axis of the complex plane.
Moreover, it consists of two qualitatively different parts:

(i) the essential spectrum, which is given by the formula
\begin{equation}
\sigma_\mathrm{ess} = \left\{ \exp\left( \int_0^T \mu(x,t) dt \right)\::\: x\in[0,2\pi] \right\}
\cup\{\mathrm{c.c.}\} 
\label{Spectrum:ess}
\end{equation}

(ii) the discrete spectrum $\sigma_\mathrm{disc}$
that consists of finitely many isolated eigenvalues $\lambda$,
which can be found using a characteristic integral equation,
as explained in~\cite[Section~4]{ome22}.

Note that if $a(x,t)$ is obtained by solving the self-consistency equation~(\ref{Eq:SC})
and hence it satisfies
$$
a(x,t/\omega) = U(x,t) = \mathcal{U}\left( W(x,t), \fr{ \eta_0 + i \gamma }{2\omega}, \omega \right),
$$
where $(W(x,t),\omega)$ is a solution of Eq.~(\ref{Eq:SC}),
then we can use Proposition~\ref{Proposition:Stability} and formula~(\ref{Def:mu}) to show
$$
\left| \exp\left( \int_0^T \mu(x,t) dt \right) \right| < 1\quad\mbox{for all}\quad x\in[0,2\pi].
$$
In this case, the essential spectrum $\sigma_\mathrm{ess}$ lies in the open unit disc~$\mathbb{D}$
and therefore it cannot contribute to any linear instability of the zero solution of Eq.~(\ref{Eq:V}).

\begin{figure}[t]
\centering
\includegraphics[scale=0.9]{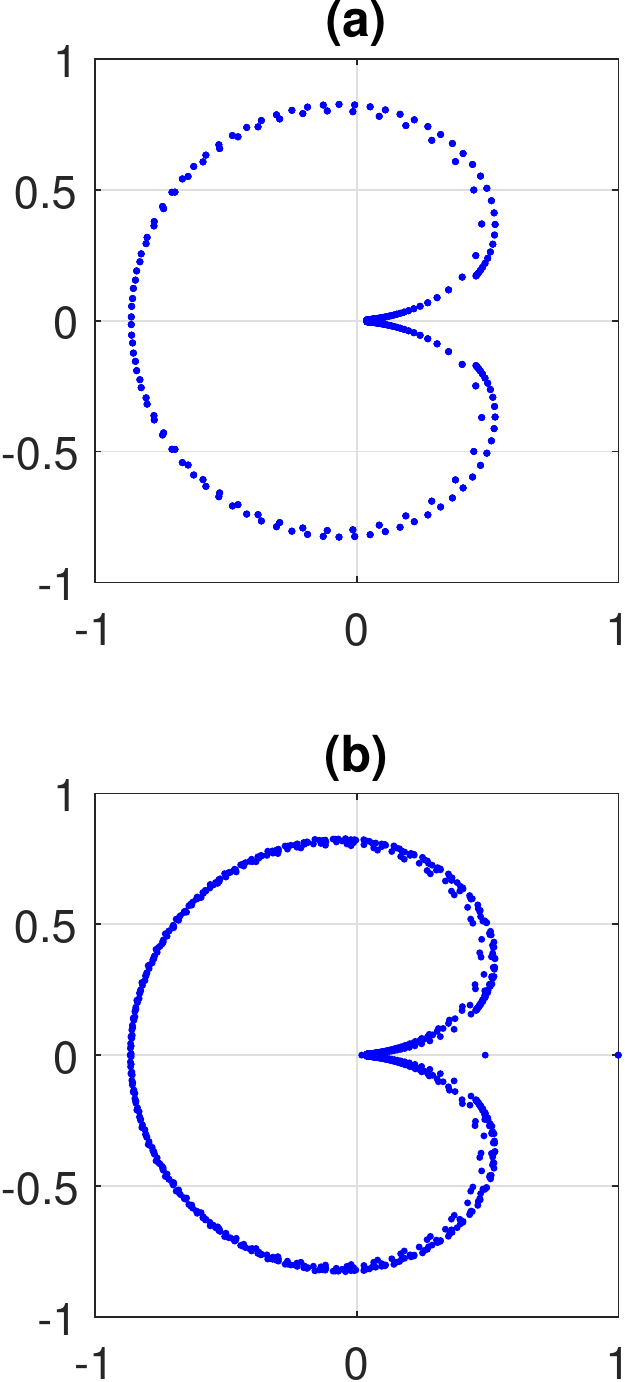}
\caption{(a) The essential spectrum given by~\eqref{Spectrum:ess} for the periodic solution
shown in Fig.~\ref{fig:symm}. (b) Floquet multipliers of the same periodic solution 
found using the technique explained at the start of Sec.~\ref{sec:periodic}.
For both calculations the spatial domain has been discretized using 512
evenly spaced points.}
\label{fig:spec}
\end{figure}

To illustrate the usefulness of formula~(\ref{Spectrum:ess}),
in Fig.~\ref{fig:spec}~(a) we plot the essential spectrum for the periodic solution
shown in Fig.~\ref{fig:symm}. In Fig.~\ref{fig:spec}~(b) we show the Floquet multipliers of the
same periodic solution, where we have found the solution and its stability in the
conventional way, of discretizing the domain and finding a periodic solution of a large
set of coupled ordinary differential equations. In panel~(b) we see several real
Floquet multipliers that do not appear in panel~(a);
these are presumably part of the  discrete spectrum.
Note that calculating the discrete spectrum by the method of~\cite[Section~4]{ome22}
is numerically difficult, so we do not do that here.

\subsection{Formula for firing rates}
\label{Sec:FiringRates}

One quantity of interest in a network of model neurons such as~\eqref{Eq:Theta}
is their firing rate. The firing rate of the $k$th neuron is defined by
$$
f_k = \fr{1}{2\pi} \left\langle \df{\theta_k}{t} \right\rangle_\mathrm{T},
$$
where the angled brackets $\langle\cdot\rangle_\mathrm{T}$ indicate a long-time average.
In the case of large $N$, we can also consider the average firing rate
\begin{equation}
f(x) = \fr{1}{\#\{ k : | x_k - x | < \pi/\sqrt{N} \}} \sum\limits_{| x_k - x | < \pi/\sqrt{N}} f_k,
\label{Def:f}
\end{equation}
where $x_k = 2\pi k/N$ is the spatial positions of the $k$th neuron
and the averaging takes place over all neurons
in the $(\pi/\sqrt{N})$-vicinity of the point $x\in[0,2\pi]$.
Note that while the individual firing rates $f_k$ are usually randomly distributed
due to the randomness of the excitability parameters $\eta_k$,
the average firing rate $f(x)$ converges to a continuous
(and even smooth) function for $N\to\infty$.
Moreover, the exact prediction of the limit function $f(x)$
can be given, using only the corresponding solution $z(x,t)$ of Eq. (\ref{Eq:L}).
To show this, we write Eq.~\eqref{Eq:Theta} as
\begin{equation}
\df{\theta_k}{t} = \mathrm{Re}\left\{ 1 - e^{i \theta_k} + ( \eta_k + \kappa I_k ) ( 1 + e^{i \theta_k} ) \right\}.
\label{Eq:Theta:I}
\end{equation}
We recall that in deriving~\eqref{Eq:L} from~\eqref{Eq:Theta}
we introduce a probability distribution $\rho(\theta,x,\eta,t)$
which satisfies a continuity equation~\cite{lai15,omewol14,lai14a}.
At a given time $t$, \linebreak $\rho(\theta,x,\eta,t)d\theta d\eta dx$ is the probability
that a neuron with a position in $[x, x + dx]$ and intrinsic drive in $[\eta, \eta + d\eta]$ 
has its phase in $[\theta, \theta + d\theta]$.
Moreover, in the case of the Lorentzian distribution of parameters~$\eta_k$,
the probability distribution $\rho(\theta,x,\eta,t)$ satisfies the relations
\begin{eqnarray}
&&
\int_{-\infty}^\infty d \eta \int_0^{2\pi} \rho(\theta,x,\eta,t) e^{i \theta} d\theta = z(x,t),
\label{rho_to_z:1}\\
&&
\int_{-\infty}^\infty d\eta \int_0^{2\pi} \eta \rho(\theta,x,\eta,t) e^{i \theta} d\theta = (\eta_0 + i \gamma) z(x,t),
\label{rho_to_z:2}
\end{eqnarray}
which are obtained by a standard contour integration in the complex plane~\cite{ottant08}.
The relation~(\ref{rho_to_z:1}) has been already used to calculate the continuum limit analog of~(\ref{eq:I})
$$
I(x,t) = \int_0^{2\pi} K(x-y) H_n(z(y,t)) dy = \mathcal{K} H_n(z).
$$
Inserting this instead of $I_k(t)$ into Eq.~(\ref{Eq:Theta:I})
and replacing the time and index averaging in~(\ref{Def:f})
with the corresponding averaging over the probability denisty, we obtain
\begin{eqnarray*}
&&
f(x) =  \lim_{\tau\to\infty}\frac{1}{2\pi\tau}\int_0^\tau\int_{-\infty}^\infty \int_0^{2\pi}
\mathrm{Re} \left\{ 1 - e^{i \theta} \right. \\[2mm]
&&
\left. + ( \eta + \kappa I(x,t) ) ( 1 + e^{i \theta} ) \right\} \rho(\theta,x,\eta,t) d\theta\ d\eta\ dt.
\end{eqnarray*}
The two inner integrals in the above formula can be simplified
using the relations~(\ref{rho_to_z:1}), (\ref{rho_to_z:2})
and the standard normalization condition for $\rho(\theta,x,\eta,t)$.
Thus we obtain
\begin{eqnarray*}
&&
f(x) =  \lim_{\tau\to\infty}\frac{1}{2\pi\tau}\int_0^\tau
\mathrm{Re} \left\{ 1 - z(x,t) \right. \\[2mm]
&&
\left. + ( \eta_0 + i \gamma + \kappa I(x,t) ) ( 1 + z(x,t) ) \right\} dt.
\end{eqnarray*}
Moreover, if $z = a(x,t)$ is a $T$-periodic solution of Eq.~(\ref{Eq:L}),
then the long-time average is the same as an average over one period.
So, in the periodic case, the continuum limit average firing rate equals
\begin{eqnarray}
&&
f(x) =  \fr{1}{2\pi T} \int_0^T \mathrm{Re} \left\{ 1 - a(x,t) \right. \nonumber \\
& & \left. + ( \eta_0 + i \gamma + \kappa I(x,t) ) ( 1 + a(x,t) ) \right\} dt.
\label{Formula:f}
\end{eqnarray}
(Note that with simple time rescaling formula~(\ref{Formula:f})
can be rewritten in terms of a $2\pi$-periodic solution
of the complex Riccati equation~\eqref{Eq:Riccati}, $u(x,t) = a(x,t/\omega)$.)
The expression~\eqref{Formula:f} is different from the
firing rate expression given in Sec.~\ref{Sec:Model}, but both are equally valid.

\begin{figure}[t]
\centering
\includegraphics[scale=0.6]{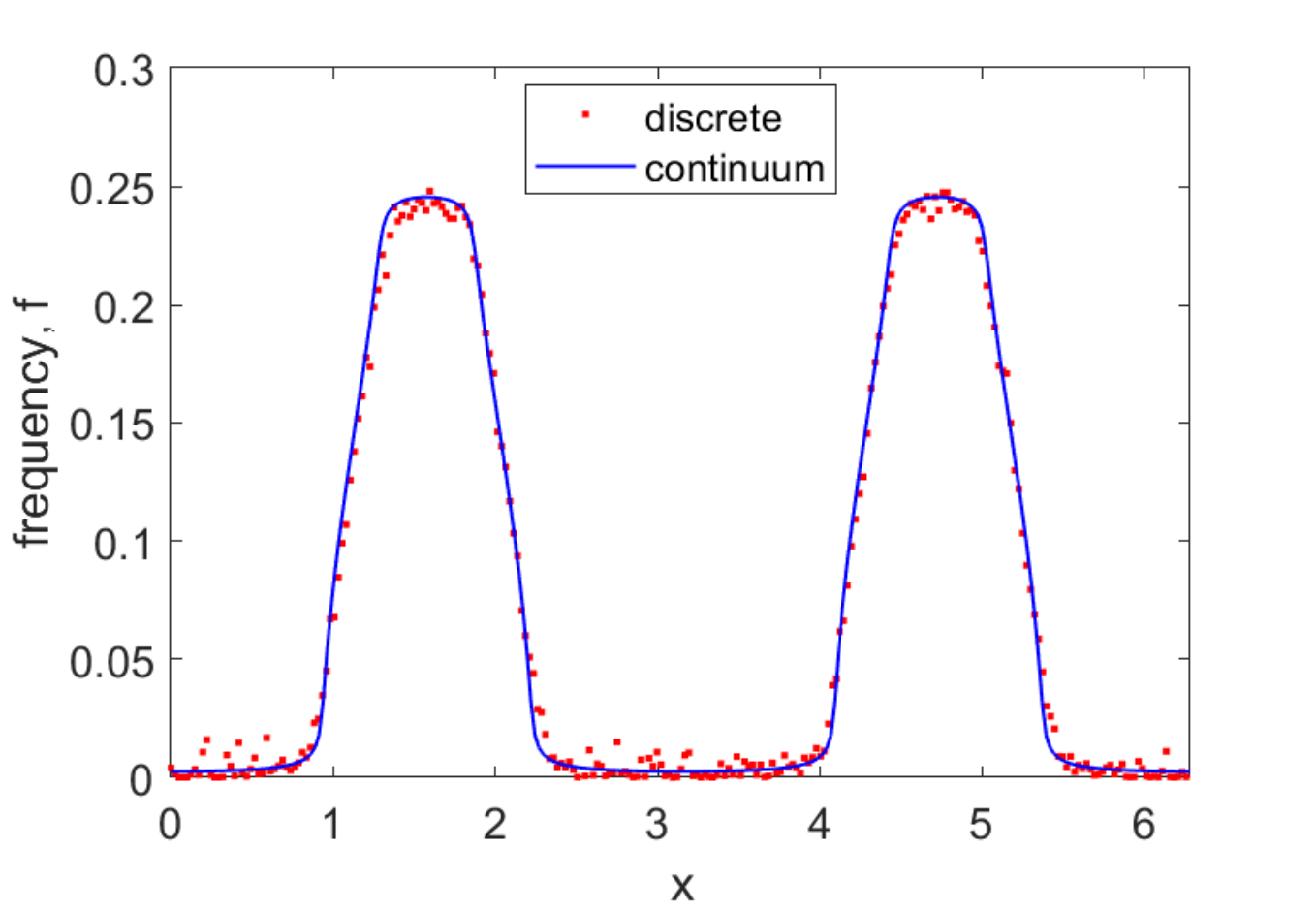}
\caption{Average firing rate for a pattern like that shown in Fig.~\ref{fig:symm}.
The curve shows $f(x)$ as given by~\eqref{Formula:f}.
The dots show values measured from direct simulations of Eq.~\eqref{Eq:Theta}.
For the discrete simulation, $N=2^{14}$ neurons were used
and the average frequency profile, $\{f_j\}, j=1,2\dots N$,
was convolved with a spatial Gaussian filter of standard deviation $0.01$ before plotting.
For clarity, not all points are shown.
Other parameters: $A=-5$, $\eta_0=-0.9$, $\kappa=1$, $\gamma=0.01$.}
\label{fig:freq}
\end{figure}

Results for a pattern like that shown in Fig.~\ref{fig:symm} are given in Fig.~\ref{fig:freq},
where we show both $f(x)$ (from~\eqref{Formula:f}) and values extracted from a long simulation
of Eq.~\eqref{Eq:Theta}. The agreement is very good.

\section{Other models}
\label{sec:other}
We now demonstrate how the approach presented above can be applied to various other
neural models.

\subsection{Delays}

Delays in neural systems are ubiquitous due to the finite velocity at which
action potentials propagate
as well as to both dendritic and synaptic processing~\cite{roxbru05,coolai09,atahut06,devmon18}.
Here we assume that all $I_j(t)$ are delayed by a fixed amount $\tau$, i.e.~we 
have Eq.~\eqref{Eq:Theta} but we replace~\eqref{eq:I} by
\begin{equation}
I_j(t) = \fr{2\pi}{N} \sum\limits_{k=1}^N K_{jk} P_n(\theta_k(t-\tau)).
\label{eq:Idel}
\end{equation}
The mean field equation is now
\begin{eqnarray}
\pf{z}{t} & = & \fr{(i \eta_0 - \gamma) ( 1 + z )^2 - i ( 1- z )^2 }{2} \nonumber \\
& + & \kappa \fr{ i ( 1 + z )^2 }{2} \mathcal{K} H_n(z(x,t-\tau)).
\label{Eq:Ldel}
\end{eqnarray}
We can write this equation in the same form as Eq.~\eqref{Eq:Riccati} but now
\begin{equation}
W(x,t) = \fr{\kappa}{2\omega} \mathcal{K} H_n(u(x,t-\omega\tau)),
\label{Def:Wdel}
\end{equation}
and the corresponding self-consistency equation is also time-delayed
$$
W(x,t) = \fr{\kappa}{2\omega} \mathcal{K} H_n\left( \mathcal{U}\left( W(x,t - \omega\tau), \fr{\eta_0 + i \gamma}{2\omega}, \omega \right) \right).
$$
We expand $W(x,t)$ as in~\eqref{W:approx}, and given $W(x,t)$,
we find the relevant $2\pi$-periodic solution of Eq.~\eqref{Eq:Riccati} as above.
The only difference comes in evaluating the projections~\eqref{Eq:vm} and~\eqref{Eq:wm}.
Instead of using $H_n(U(x,t))$ in the scalar products, we need to use $H_n(U(x,t-\omega\tau))$.

Since $U(x,t)$ is $2\pi$-periodic in time, we can evaluate it at any time using just its
values for $t\in[0,2\pi]$. Specifically,
\begin{equation}
U(x,t-\omega\tau) =
\begin{cases}
U(x,2\pi+t-\omega\tau), & 0\leq t\leq \omega\tau, \\[1mm]
U(x,t-\omega\tau), & \omega\tau< t\leq 2\pi.
\end{cases}
\end{equation}
Note that this approach would also be applicable if one had a distribution of 
delays~\cite{leeott09,lailon03} or even state-dependent delays~\cite{keakra19}.

\begin{figure}[t]
\centering
\includegraphics[scale=0.6]{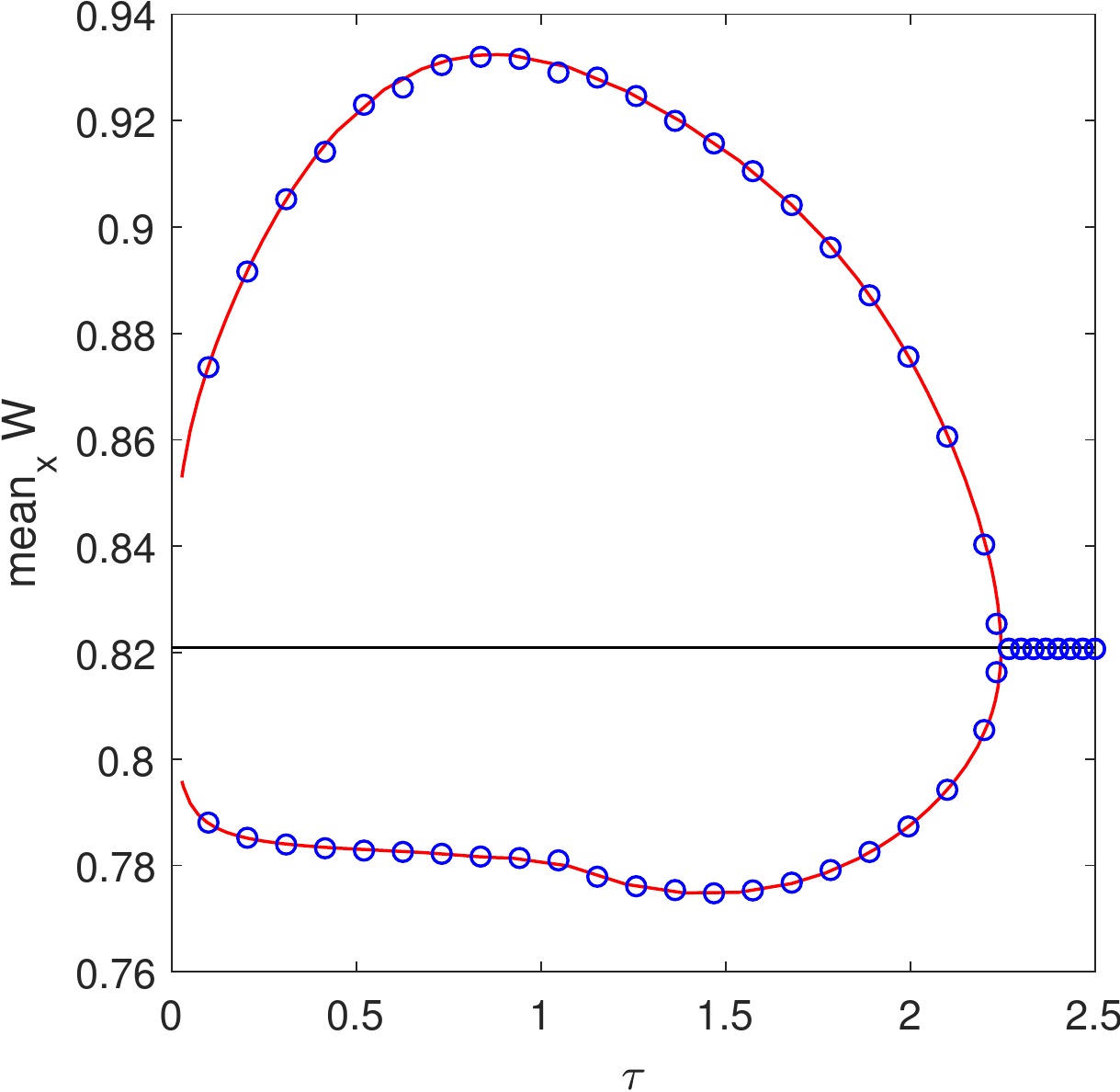}
\caption{The vertical axis relates to averaging $W(x,t)$ over $x$. For a periodic
solution, the maximum and minimum over one period of this quantity is plotted.
The black horizontal line corresponds to the steady state which is stable at $\tau=2.5$.
Circles: measured from direct simulations of Eq.~\eqref{Eq:Ldel}.
Other parameters: $A=-5$, $\eta_0=-2$, $\kappa=1$, $\gamma=0.1$.}
\label{fig:delayB}
\end{figure}

As an example, we show in Fig.~\ref{fig:delayB} the results of varying the delay
$\tau$ on a solution of the form shown in Fig.~\ref{fig:asymmA}. Increasing $\tau$
leads to the destruction of the periodic solution in an apparent supercritical Hopf
bifurcation.

\subsection{Two populations}

Neurons fall into two major categories: excitatory and inhibitory. A model consisting of a single
population with a coupling function of the form~\eqref{Coupling:Cos} is often used
as an approximation to a two-population model~\cite{esnrox17}. 
Here we consider a two population model which
supports a travelling wave. The mean field equations are

\begin{eqnarray}
\pf{u}{t} & = & \fr{(i \eta_u - \gamma) ( 1 + u )^2 - i ( 1- u )^2 }{2} \nonumber \\
& + &  \fr{ i ( 1 + u )^2 }{2} \left[w_\mathrm{ee}\mathcal{K} H_n(u)-w_\mathrm{ei}\mathcal{K} H_n(v)\right], \label{eq:twoA} \\
\pf{v}{t} & = & \fr{(i \eta_v - \gamma) ( 1 + v )^2 - i ( 1- v )^2 }{2} \nonumber \\
& + &  \fr{ i ( 1 + v )^2 }{2} \left[w_\mathrm{ie}\mathcal{K} H_n(u)-w_\mathrm{ii}\mathcal{K} H_n(v)\right] \label{eq:twoB}
\end{eqnarray}
where $u(x,t)$ is the complex-valued order parameter for the excitatory population and
$v(x,t)$ is that for the inhibitory population. The non-negative connectivity kernel
between and within populations is the same:
\[
   K(x)=\frac{1}{2\pi}(1+\cos{x})
\]
and there are four connection strengths within and between populations:
$w_\mathrm{ee}$, $w_\mathrm{ei}$, $w_\mathrm{ie}$ and $w_\mathrm{ii}$.
Similar models have been studied in~\cite{blowyl05,pinerm01}.

\begin{figure}[t!]
\centering
\vspace{10mm}
\includegraphics[scale=0.75]{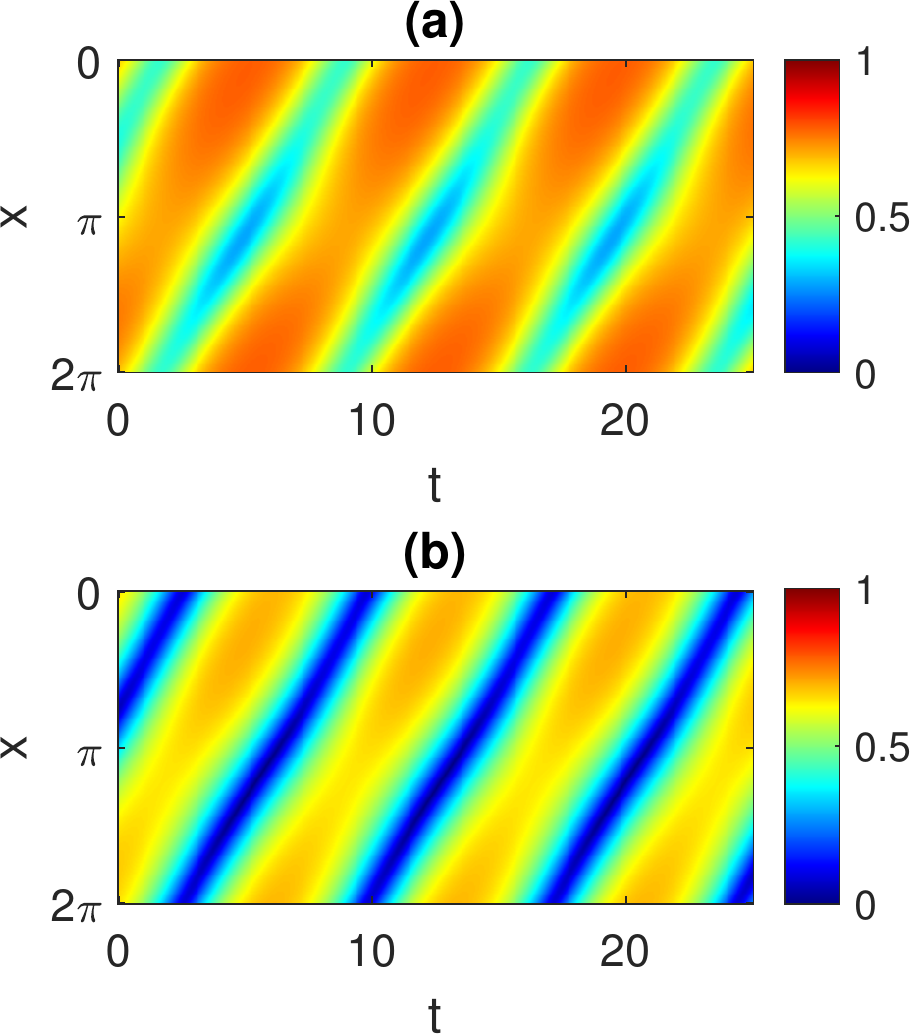}
\caption{A modulated travelling wave solution of Eqs.~\eqref{eq:twoA}--\eqref{eq:twoB}.
(a): $|u|$; (b): $|v|$.
Other parameters: $\eta_0=0.1$, $\eta_v=0.1$, $\epsilon=0.01$, $\gamma=0.03$,
$w_\mathrm{ee}=1$, $w_\mathrm{ei}=0.7$, $w_\mathrm{ie}=0.3$, $w_\mathrm{ii}=0.1$.}
\label{fig:twopop}
\end{figure}

For some parameter values, such a system supports a travelling wave with a constant profile.
Such a wave can be found very efficiently using the techniques discussed here, and that was
done for a travelling chimera in~\cite{ome23}. However, here we consider a slightly
different case: that where the mean drive to the excitatory population, $\eta_u$, is
spatially modulated. We thus write 
\[
   \eta_u = \eta_0 + \epsilon\sin{x}.
\]
For small $|\epsilon|$ the travelling wave persists, but not with a constant profile.
An example is shown in Fig.~\ref{fig:twopop}. Note that such a solution is periodic in time.

By rescaling time we can write Eqs.~\eqref{eq:twoA}--\eqref{eq:twoB} as
\begin{eqnarray}
\pf{\tilde{u}}{t} &=& i \left( w_\mathrm{ee}W_u-w_\mathrm{ei}W_v + \zeta_u - \fr{1}{2\omega} \right) \nonumber \\
& + & 2 i \left( w_\mathrm{ee}W_u-w_\mathrm{ei}W_v + \zeta_u + \fr{1}{2\omega} \right) \tilde{u} \nonumber\\
&+& i \left(w_\mathrm{ee}W_u-w_\mathrm{ei}W_v + \zeta_u - \fr{1}{2\omega} \right) \tilde{u}^2, \\
\pf{\tilde{v}}{t} &=& i \left( w_\mathrm{ie}W_u-w_\mathrm{ii}W_v + \zeta_v - \fr{1}{2\omega} \right) \nonumber \\
& + & 2 i \left( w_\mathrm{ie}W_u-w_\mathrm{ii}W_v + \zeta_v + \fr{1}{2\omega} \right) \tilde{v} \nonumber\\
&+& i \left(w_\mathrm{ie}W_u-w_\mathrm{ii}W_v + \zeta_v - \fr{1}{2\omega} \right) \tilde{v}^2,
\end{eqnarray}
where $\tilde{u}(x,t)\equiv u(x,t/\omega)$, $\tilde{v}(x,t)\equiv v(x,t/\omega)$,
\begin{eqnarray}
W_u(x,t) & = &  \fr{1}{2\omega} \mathcal{K} H_n(u), \\
W_v(x,t) & = &  \fr{1}{2\omega} \mathcal{K} H_n(v)
\end{eqnarray}
and
$$
\zeta_u = \fr{\eta_u + i \gamma}{2\omega} = \fr{\eta_0 + \epsilon\sin{x} + i \gamma}{2\omega}, \qquad \zeta_v = \fr{\eta_v + i \gamma}{2\omega}.
$$
In the same way as above, we can derive self-consistency equations of the
form~\eqref{Eq:SC} for $W_u(x,t)$ and $W_v(x,t)$.
Doing so, we obtain a system of two coupled equations
\begin{eqnarray*}
W_u(x,t) & = & \fr{\kappa}{2\omega} \mathcal{K} H_n\bigg( \mathcal{U}\bigg( w_\mathrm{ee} W_u(x,t)%
- w_\mathrm{ei}W_v(x,t),  \nonumber \\
& & \left.\left.\fr{\eta_0 + \epsilon\sin{x} + i \gamma}{2\omega}, \omega \right) \right),\\[2mm]
W_u(x,t) & = & \fr{\kappa}{2\omega} \mathcal{K} H_n\bigg( \mathcal{U}\bigg( w_\mathrm{ie} W_u(x,t) - w_\mathrm{ii}W_v(x,t), \nonumber \\
& & \left. \left. \fr{\eta_v + i \gamma}{2\omega}, \omega \right) \right).
\end{eqnarray*}

One difference between this model and the ones studied above is
that the solution cannot be shifted by a constant amount in $x$ to ensure
that it is always even (or odd) about a particular point in the domain. Thus we need to write
\begin{eqnarray}
W_u(x,t) & = & \sum\limits_{m=0}^{2F} ( v_m^u + w_m^u \sin x +z_m^u \cos x ) \psi_m(t) \\
W_v(x,t) & = & \sum\limits_{m=0}^{2F} ( v_m^v + w_m^v \sin x +z_m^v \cos x ) \psi_m(t)
\end{eqnarray}
These equations contain $6(2F+1)$ unknowns and we find them (and $\omega$) in the same way as
above, by projecting the self-consistency equations for $W_u(x,t)$ and $W_v(x,t)$ onto the
different spatio-temporal Fourier modes to obtain equations similar to~\eqref{Eq:vm}--\eqref{Eq:wm}.

\begin{figure}[t!]
\centering
\includegraphics[scale=0.6]{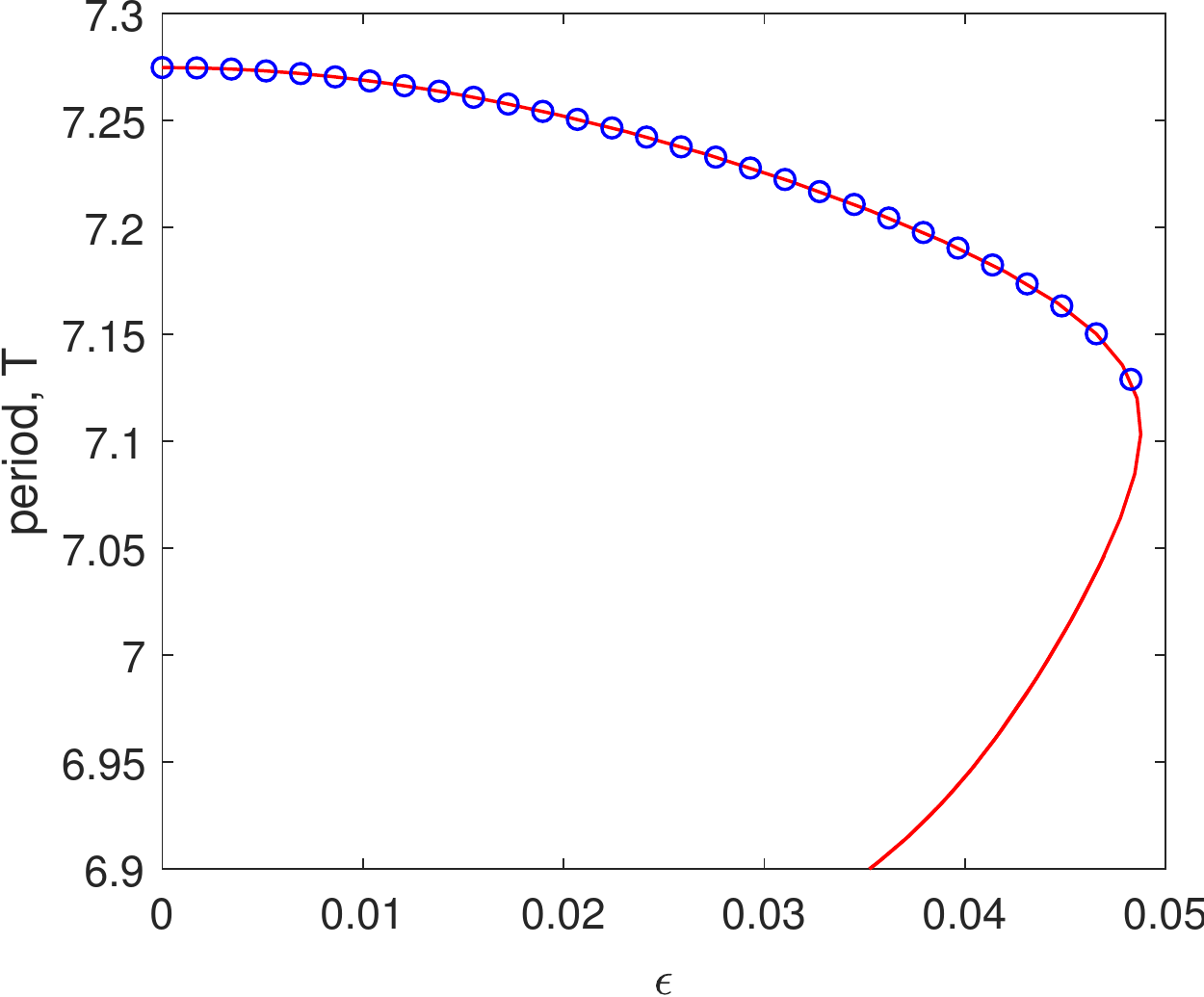}
\caption{Period, $T$, of a modulated travelling wave solution of Eqs.~\eqref{eq:twoA}--\eqref{eq:twoB}
as a function of heterogeneity strength $\epsilon$.
Circles are from direct simulation of Eqs.~\eqref{eq:twoA}--\eqref{eq:twoB}.
Other parameters are as in Fig.~\ref{fig:twopop}.}
\label{fig:varyep}
\end{figure}

The results of varying the heterogeneity strength $\epsilon$ are shown in Fig.~\ref{fig:varyep}.
Increasing heterogeneity decreases the period of oscillation, and eventually the travelling
wave appears to be destroyed in a saddle-node bifurcation.

We conclude this section by noting that for some parameter values
the model~\eqref{eq:twoA}--\eqref{eq:twoB} can show periodic solutions
which do not travel, like those shown in Sec.~\ref{sec:periodic}.
Likewise, the model in Sec.~\ref{Sec:Model} can support travelling waves for $\kappa=2$.

\subsection{Winfree oscillators}

One of the first models of interacting oscillators studied is the Winfree 
model~\cite{aristr01,laibla21,pazmon14,galmon17}. We consider a spatially-extended network
of Winfree oscillators whose dynamics are given by
\[
  \df{\theta_j}{t}=\omega_j+\epsilon\frac{2\pi Q(\theta_j)}{N}\sum_{k=1}^N K_{jk}P(\theta_k)
\]
where $K_{jk}=K(2\pi|j-k|/N)$ for some $2\pi$-periodic coupling function $K$, 
$Q(\theta)=-\sin{\theta}/\sqrt{\pi}$ and
$P(\theta)=(2/3)(1+\cos{\theta})^2$ is a pulsatile function with its peak at $\theta=0$.
The $\omega_j$ are randomly chosen from a Lorentzian with centre $\omega_0$ and
width $\Delta$ and $\epsilon$ is the overall coupling strength.

\begin{figure}[t!]
\centering
\includegraphics[scale=0.75]{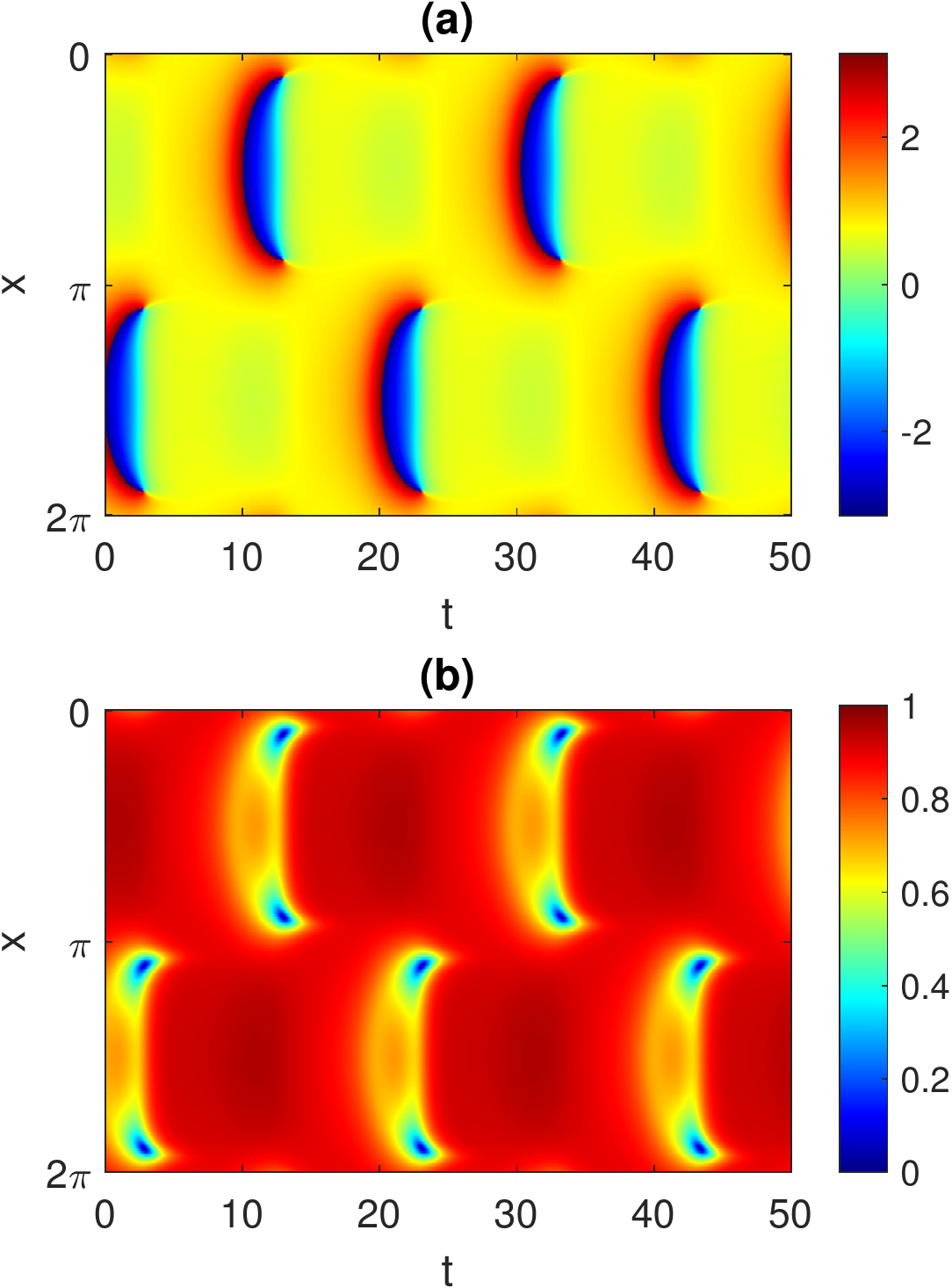}
\caption{A periodic solution of Eq.~\eqref{eq:win}. (a): $\arg{(z(x,t))}$. (b): $|z(x,t)|$.
Parameters: $\omega_0=1$, $\Delta=0.1$, $\epsilon=2.1$.}
\label{fig:winexam}
\end{figure}

In the limit $N\to\infty$, using the Ott/Antonsen ansatz,
one finds that the network is described by the equation~\cite{lai17}
\begin{equation}
  \pf{z}{t}=\frac{\epsilon}{2\sqrt{\pi}} \mathcal{K} \widehat{H}(z) +(i\omega_0-\Delta)z-\frac{\epsilon}{2\sqrt{\pi}} z^2 \mathcal{K} \widehat{H}(z),
\label{eq:win}
\end{equation}
where the integral operator $\mathcal{K}$ is again defined by~(\ref{Def:K})
and
\[
   \widehat{H}(z)=(2/3)[3/2+z+\bar{z}+(z^2+\bar{z}^2)/4].
\]
A typical periodic solution of Eq.~(\ref{eq:win}) for the choice
$$
K(x)=0.1+0.3\cos{x},
$$
is shown in Fig.~\ref{fig:winexam}.

If we rescale time by the frequency of periodic solution $\omega>0$,
defining $u(x,t) = z(x,t/\omega)$, and denote
$$
W(x,t) = \frac{\epsilon}{2\sqrt{\pi}\omega} \mathcal{K} \widehat{H}(u),
$$
then Eq.~(\ref{eq:win}) can be recast as a complex Riccati equation
\begin{equation}
\pf{u}{t} = W(x,t) + i \fr{\omega_0 + i \Delta}{\omega} u - W(x,t) u^2.
\label{Eq:Riccati:win}
\end{equation}
From the Proposition~2 in~\cite{ome23}, it follows that for every $\omega,\Delta > 0$,
$\omega_0\in\mathbb{R}$ and for every real-valued, $2\pi$-periodic in $t$ function $W(x,t)$,
Eq.~(\ref{Eq:Riccati:win}) has a $2\pi$-periodic solution lying in the open unit disc $\mathbb{D}$.
Denoting the corresponding solution operator
$$
\widehat{\mathcal{U}}\::\: C_\mathrm{per}([0,2\pi];\mathbb{R})\times\mathbb{C}_\mathrm{up}\to C_\mathrm{per}([0,2\pi];\mathbb{D}),
$$
we easily obtain a self-consistency equation for periodic solutions of Eq.~(\ref{Eq:Riccati:win})
\begin{equation}
W(x,t) = \frac{\epsilon}{2\sqrt{\pi}\omega} \mathcal{K} \widehat{H}\left(  \widehat{\mathcal{U}}\left( W(x,t), \fr{\omega_0 + i \Delta}{\omega} \right) \right).
\label{Eq:SC:win}
\end{equation}
Since the Poincar{\'e} map of Eq.~(\ref{Eq:Riccati:win}) coincides with the M{\"o}bius transformation,
we can again use the calculation scheme of Sec.~\ref{Sec:Moebius} to find the value of operator~$\widehat{\mathcal{U}}$.
Thus we can solve Eq.~(\ref{Eq:SC:win}) numerically for the real-valued field $W(x,t)$ and frequency $\omega$.

\begin{figure}[t!]
\centering
\includegraphics[scale=0.7]{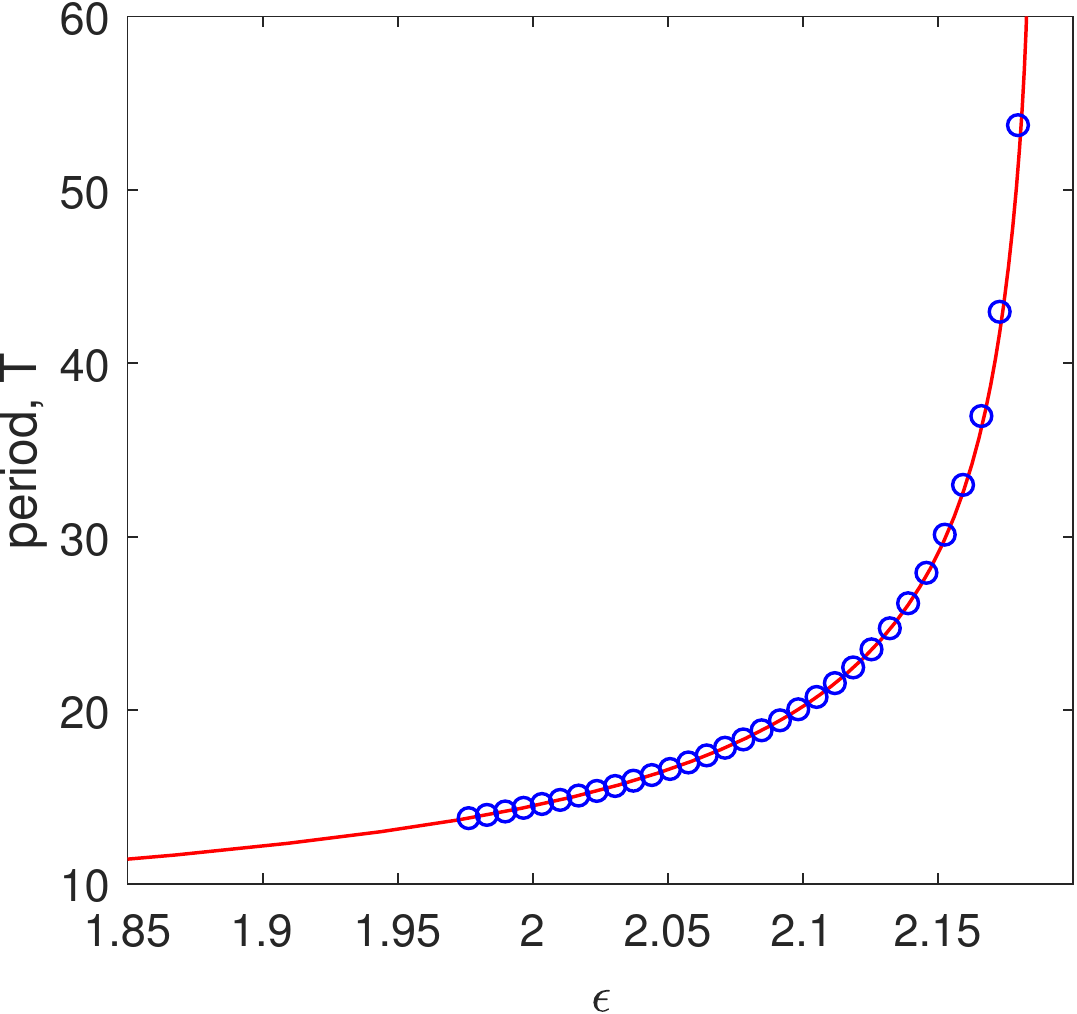}
\caption{Period, $T$, of periodic solutions of Eq.~\eqref{eq:win}
of the form shown in Fig.~\ref{fig:winexam}.
Circles are from direct simulations of Eq.~\eqref{eq:win}.
Parameters: $\omega_0=1$, $\Delta=0.1$.}
\label{fig:winper}
\end{figure}

Following the periodic solution shown in Fig.~\ref{fig:winexam}
as $\epsilon$ is varied we obtain Fig.~\ref{fig:winper}.
As shown by the circles (from direct simulations of Eq.~\eqref{eq:win})
this solution is not always stable.
As $\epsilon$ is decreased the solution loses stability to a uniformly travelling wave,
and the period of this wave is not plotted.

\section{Discussion}
\label{sec:disc}

We considered time-periodic solutions of the Eqs.~\eqref{Eq:L}--\eqref{Def:K},
which exactly describe the asymptotic dynamics of the network~\eqref{Eq:Theta} in the limit
of $N\to\infty$. At every point in space, Eq.~\eqref{Eq:L} is a Riccati equation
and we used this to derive a self-consistency equation
that every periodic solution of Eq.~\eqref{Eq:L} must satisfy.
The Poincar{\'e} map of the Riccati equation is a M{\"o}bius map and we can determine
this map at every point in space using just three numerical solutions of the Riccati equation.
Knowing the M{\"o}bius map enables us to numerically solve the self-consistency equation in a 
computationally efficient manner. We showed the results of numerically continuing several
types of periodic solutions as a parameter was varied.

We derived equations governing the stability of such periodic solutions, but solving these
equations is numerically challenging. We also derived the expression for the mean firing rate
of neurons in a network in terms of the quantities already calculated in the self-consistency
equation. We finished in Sec.~\ref{sec:other} by demonstrating the application of our approach
to several other models involving delays, two populations of neurons, and a network of Winfree
oscillators.

Our approach relies critically on the mathematical form of the continuum-level equations
(they can be written as a Riccati equation) which are derived using the Ott/Antonsen ansatz,
valid only for phase oscillators whose dynamics and coupling involve sinusoidal
functions of phases or phase differences. Other systems for which our approach should
be applicable include two-dimensional networks which support moving or ``breathing''
solutions~\cite{batcle21}; however the coupling function would have to be of the
form such that the integral equivalent to~\eqref{Def:K} could be written exactly using
a small set of spatial basis functions. Another application would be to any system which is
periodically forced in time and responds in a periodic way~\cite{segbi20,reyhug22,schavi18}.

\section*{Appendix}


Let us consider a complex Riccati equation
\begin{equation}
\df{z}{t} = c_0(t) + c_1(t) z + c_2(t) z^2
\label{Eq:Riccati:A}
\end{equation}
with $2\pi$-periodic complex-valued coefficients $c_0(t)$, $c_1(t)$ and $c_2(t)$.
In this section we prove a statement
which is a modified version of Proposition~3 from~\cite{ome23}.

\begin{proposition}
Suppose that there is $c_* > 0$ such that
\begin{equation}
\Real( \overline{c_0(t)} z + c_1(t) + c_2(t) z ) \le - c_* \Real(z + 1)
\label{Ineq:c_star}
\end{equation}
for all $z\in\overline{\mathbb{D}}$ and $0\le t \le 2\pi$,
and that $z = - 1$ is not a fixed point of Eq.~(\ref{Eq:Riccati:A}).
Then, the Poincar{\'e} map of Eq.~(\ref{Eq:Riccati:A}) is described
by a hyperbolic or loxodromic M{\"o}bius transformation.
Moreover, the stable fixed point of this map lies in the open unit disc~$\mathbb{D}$,
while the unstable fixed point lies in the complementary domain $\hat{\mathbb{C}}\backslash\overline{\mathbb{D}}$,
where $\hat{\mathbb{C}} = \mathbb{C}\cup\{\infty\}$ is the extended complex plane.
For Eq.~(\ref{Eq:Riccati:A}) this means that it has exactly one stable $2\pi$-periodic solution
and this solution satisfies $|z(t)| < 1$ for all $0\le t \le 2\pi$.
\label{Proposition:Sln:Theta}
\end{proposition}

{\bf Proof:} The fact that the Poincar{\'e} map of Eq.~(\ref{Eq:Riccati:A})
is described by a M{\"o}bius transformation was shown elsewhere~\cite{cam97,wil08}.
So we only need to show that such a M{\"o}bius transformation $\mathcal{M}(z)$
maps all points of the unit circle $|z| = 1$ into the open unit disc $\mathbb{D}$ (but not on its boundary).
Then we can repeat the arguments of the proof of Proposition~2 from~\cite{ome23}.

Suppose the opposite. Then Eq.~(\ref{Eq:Riccati:A}) has a solution $z_*(t)$ such that $|z_*(0)| = |z_*(2\pi)| = 1$.
Due to the inequality~(\ref{Ineq:c_star}) and the Proposition~1 from~\cite{ome23}
this solution satisfies $|z_*(t)| \le 1$ for all $t\in[0,2\pi]$.
Moreover, since $z = -1$ is not a fixed point of Eq.~(\ref{Eq:Riccati:A}),
we can always choose $t_*\in(0,2\pi)$ such that $z_*(t)\ne -1$ for $t\in[t_*,2\pi)$.
Then the Mean Value Theorem implies
\begin{equation}
0 \le |z_*(2\pi)|^2 - |z_*(t_*)|^2 = 2 \pi \df{|z_*|^2}{t}(t_{**})
\label{Ineq:z_star}
\end{equation}
for some $t_{**}\in(t_*,2\pi)$.
On the other hand, due to our assumptions, we have
\begin{eqnarray*}
\df{|z_*|^2}{t} &=& 2 \Real\left( \overline{c_0(t)} z_*(t) + c_1(t) + c_2(t) z_*(t) \right) \\[2mm]
&\le& - 2 c_* \Real(z_*(t) + 1)
\end{eqnarray*}
and therefore
$$
\df{|z_*|^2}{t}(t_{**}) \le - 2 c_* \Real(z_*(t_{**}) + 1) < 0.
$$
This is a contradiction to~(\ref{Ineq:z_star}), which completes the proof.~\qed

\begin{remark}
If the conditions of Proposition~\ref{Proposition:Sln:Theta} are fulfilled,
then the stable solution of Eq.~(\ref{Eq:Riccati:A}) is also asymptotically stable.
This result follows from the properties of stable fixed points
of hyperbolic and loxodromic M{\"o}bius transformations.
\label{Remark:AsymptoticStability}
\end{remark}

\begin{remark}
For every fixed $x$ the equation~(\ref{Eq:Riccati}) from the main text
is equivalent to Eq.~(\ref{Eq:Riccati:A}) with
$$
c_0(t) = c_2(t) = i \left( W(x,t) + \zeta - \fr{1}{2\omega} \right)
$$
and
$$
c_1(t) = 2 i \left( W(x,t) + \zeta + \fr{1}{2\omega} \right),
$$
where $W(x,t)$ is a real-valued function, $\omega$ is a positive constant,
and $\zeta = (\eta_0 + i \gamma) / (2\omega)$
with $\eta_0\in\mathbb{R}$ and $\gamma > 0$.
In this case, we have
$$
\Real( \overline{c_0(t)} z + c_1(t) + c_2(t) z ) = - \fr{\gamma}{\omega} \Real(z + 1)
$$
and therefore inequality~(\ref{Ineq:c_star}) is satisfied.
On the other hand, we have
$$
c_0(t) + c_1(t) (-1) + c_2(t) (-1)^2 = - 2 i / \omega \ne 0
$$
and therefore $z = -1$ is not a fixed point of the corresponding Riccati equation.
Thus, all of the conclusions of Proposition~\ref{Proposition:Sln:Theta} hold true for Eq.~(\ref{Eq:Riccati}).
\label{Remark:Sln:Theta}
\end{remark}

\begin{remark}
If the conditions of Proposition~\ref{Proposition:Sln:Theta} are fulfilled,
then the stable solution of Eq.~(\ref{Eq:Riccati:A}) can be computed in the following way.
\smallskip

(i) One solves Eq.~(\ref{Eq:Riccati:A}) on the interval $t\in(0,2\pi]$
with three different initial conditions $z(0) = z_k\in\mathbb{D}$, $k=1,2,3$,
and obtains three solutions $Z_k(t)$.
Since each $z_k$ lies in the open unit disc~$\mathbb{D}$
this automatically implies $|Z_k(t)| < 1$ for all $t\in(0,2\pi]$.
\smallskip

(ii) One denotes $w_k = Z_k(2\pi)$. Then, due to the properties
of Poincar{\'e} map one has $w_k = \mathcal{M}(z_k)$, $k=1,2,3$,
where $\mathcal{M}(z)$ is a M{\"o}bius transformation representing this map.
The above three relations can be used to reconstruct the map $\mathcal{M}(z)$, namely
$$
\mathcal{M}(z) = \frac{a z + b}{c z + d}
$$
where
\begin{eqnarray*}
&&
a = \det \left(
\begin{array}{ccc}
 z_1 w_1 & w_1 & 1 \\[2mm]
 z_2 w_2 & w_2 & 1 \\[2mm]
 z_3 w_3 & w_3 & 1
\end{array}
\right),
\quad
b = \det \left(
\begin{array}{ccc}
 z_1 w_1 & z_1 & w_1 \\[2mm]
 z_2 w_2 & z_2 & w_2 \\[2mm]
 z_3 w_3 & z_3 & w_3
\end{array}
\right),
\\[2mm]
&&
c = \det \left(
\begin{array}{ccc}
 z_1 & w_1 & 1 \\[2mm]
 z_2 & w_2 & 1 \\[2mm]
 z_3 & w_3 & 1
\end{array}
\right),
\quad\phantom{w_1}
d = \det \left(
\begin{array}{ccc}
 z_1 w_1 & z_1 & 1 \\[2mm]
 z_2 w_2 & z_2 & 1 \\[2mm]
 z_3 w_3 & z_3 & 1
\end{array}
\right).
\end{eqnarray*}
\smallskip

(iii) Once the map $\mathcal{M}(z)$ is known, one can find its fixed points by
solving the quadratic equation
$$
c z^2 + d z - a z - b = 0.
$$
This yields two roots
$$
z_\pm = \fr{a - d \pm \sqrt{ (a - d)^2 + 4 b c }}{2 c}.
$$
\smallskip

(iv) Choosing from the roots $z_+$ and $z_-$ the one
that lies in the unit disc $\mathbb{D}$,
one obtains the initial condition that determines the periodic solution of interest.
The latter can be computed by solving Eq.~(\ref{Eq:Riccati:A}) with this initial condition.
\smallskip

(v) Sometime it may happen that the Poincar{\'e} map $\mathcal{M}(z)$
is strongly contracting so that
$$
|w_1 - w_2 | + |w_3 - w_2 | < 10^{-8},
$$
where the value $10^{-8}$ is chosen through experience.
In this case, the calculations in steps~(ii) and~(iii) become inaccurate.
Then the initial condition of the periodic solution of interest
is approximately given by the average $(w_1 + w_2 + w_3) /3$.
\smallskip

The above steps (i)--(v) can be understood as a constructive definition
of the solution operator of Eq.~(\ref{Eq:Riccati:A}),
which for every admissible choise of $2\pi$-periodic coefficients
$c_1(t)$, $c_2(t)$ and $c_3(t)$ yields the corresponding
stable $2\pi$-periodic solution of Eq.~(\ref{Eq:Riccati:A}).
More detailed justification of this definition can be found in~\cite[Section~4]{ome23}.
The algorithm in Sec.~\ref{Sec:Moebius} consists of applying the procedure above at every
point on the spatial grid (in parallel).

\label{Remark:Sln:Operator}
\end{remark}


\section*{Acknowledgements}
The work of O.E.O. was supported by the Deutsche Forschungsgemeinschaft
under Grant No. OM 99/2-2.

%
\section*{Declarations}
Competing interests: The authors have no competing interests to declare
that are relevant to the content of this article. 


Code availability: Code for calculating any of the results presented here
is available on reasonable request from the authors.

\section*{Author contributions}

\paragraph*{Conceptualization:}\: C.R.~Laing, O.E.~Omel'chenko.

\paragraph*{Formal analysis:}\: C.R.~Laing, O.E.~Omel'chenko.

\paragraph*{Funding acquisition:}\: O.E.~Omel'chenko.

\paragraph*{Investigation:}\: C.R.~Laing, O.E.~Omel'chenko.

\paragraph*{Methodology:}\: C.R.~Laing, O.E.~Omel'chenko.

\paragraph*{Software:}\: C.R.~Laing.

\paragraph*{Validation:}\: C.R.~Laing, O.E.~Omel'chenko.

\paragraph*{Visualization:}\: C.R.~Laing.

\paragraph*{Writing -- original draft:}\: C.R.~Laing, O.E.~Omel'chenko.

\paragraph*{Writing -- review \& editing:}\: C.R.~Laing, O.E.~Omel'chenko.



\end{document}